\documentclass{aa}
\usepackage{graphics}
\input{epsf.sty}

\begin{document}

   \thesaurus{04        
              (
               10.11.1; 
               10.19.1; 
               10.19.3; 
               08.05.1; 
               08.11.1  
              )
              }

    \title{Kinematics of young stars (I): Local irregularities
    \thanks{Based on data from the Hipparcos astrometry satellite
            (European Space Agency)}}

   \author{J. Torra
      \and D. Fern\'andez
      \and F. Figueras
          }

   \offprints{J. Torra}
   \mail{jordi@am.ub.es}

   \institute{Departament d'Astronomia i Meteorologia, Universitat de
              Barcelona, Av. Diagonal 647, E-08028 Barcelona, Spain
             }

   \date{Received <date> / Accepted <date>}

   \maketitle

   \begin{abstract}

The local velocity field of young stars is dominated by the galactic
rotation, the kinematics of the Gould Belt and the nearest OB associations
and open clusters, and the kinematics of the spiral structure. We
re-examined here this local velocity field by using a large sample of
nearby O and B stars from the Hipparcos Catalogue. The high quality
astrometric data are complemented with a careful compilation of radial
velocities and Str\"omgren photometry, which allows individual photometric
distances and ages to be derived. The Gould Belt extends up to 600 pc from
the Sun with an inclination with respect to the galactic plane of
$i_{\mathrm{G}} = 16$-$22\degr$ and the ascending node placed at
$\Omega_{\mathrm{G}} = 275$-$295\degr$. Approximately 60\% of the stars
younger than 60 Myr belong to this structure. The values found for the
Oort constants when different samples selected by age or distance were
used allowed us to interpret the systematic trends observed as signatures
induced by the kinematic behaviour of the Gould Belt. The contribution of
Sco-Cen and Ori OB1 complexes in the characterization of the expansion of
the Gould Belt system is also discussed. We found that a positive $K$-term
remains when these aggregates are excluded. From the kinematic behaviour
of the stars and their spatial distribution we derive an age for the Gould
Belt system in the interval 30-60 Myr.


   \end{abstract}

%
\section{Introduction}

The kinematic study of the local system of young stars offers an excellent
opportunity for understanding the history of recent star formation and for
improving our knowledge of the dynamics involved in the evolution of our
galaxy.

The motion of young nearby stars deviates considerably from the general
field of galactic rotation (see e.g. du Mont \cite{du Mont}; Clube
\cite{Clube}). The presence of a positive $K$-term, corresponding to an
overall expansion of the local system of the earliest stars was long ago
(Campbell \cite{Campbell}) recognized as the main kinematic characteristic
of the Gould Belt. This system is recognized to be a great circle inclined
some $20\degr$ respect to the galactic equator and is traced by young
stars and OB associations, HI, molecular clouds and dust. A detailed
review of the structure, kinematics and origin of the Gould Belt has
recently been undertaken by P\"oppel (\cite{Poppel}). Lesh (\cite{Lesh}),
Westin (\cite{Westin}) and Comer\'on et al. (\cite{Comeron et al.3};
hereafter referred to as CTG) found, in addition to the positive
$K$-value, departures from the values of other Oort constants when working
with stars associated with the Gould Belt. In addition to the peculiar
kinematics of the Gould Belt system, deviations from circular motion have
also been analysed in the context of the kinematic effects of the spiral
structure (Cr\'ez\'e \& Mennessier \cite{Creze et al.}; Lindblad
\cite{Lindblad}; Byl \& Ovenden \cite{Byl et al.}; Westin \cite{Westin};
Comer\'on \& Torra \cite{Comeron et al.1}, among others).

The publication of the Hipparcos data (ESA \cite{ESA2}) has made it
possible to re-analyse the galactic velocity field and its local
irregularities. Feast \& Whitelock (\cite{Feast et al.1}) re-determined
the mean values of the Oort constants from Hipparcos proper motion of
Cepheids ($A = 14.8 \pm 0.8$ km s$^{-1}$ kpc$^{-1}$, $B = -12.4 \pm 0.6$
km s$^{-1}$ kpc$^{-1}$). Feast et al. (\cite{Feast et al.2}) obtained $A =
15.1 \pm 0.3$ km s$^{-1}$ kpc$^{-1}$ by using radial velocities and the
new zero-point period-luminosity relation from Hipparcos trigonometric
parallaxes. Lindblad et al. (\cite{Lindblad et al.2}) and Torra et al.
(\cite{Torra et al.}) presented the first results concerning the structure
and kinematics of the local system of young stars. Whereas Lindblad et al.
(\cite{Lindblad et al.2}) do not rule out the possibility that the Gould
Belt is just a random configuration of two or three dominating
associations, Sterzik et al. (\cite{Sterzik et al.}) and Guillout et al.
(\cite{Guillout et al.2}), from a RASS-Tycho sample, suggest that the Belt
is a disk-like rather than a ring-like structure. The Hipparcos census of
nearby OB associations (de Zeeuw et al. \cite{de Zeeuw et al.}) and the
review of the mean astrometric parameters of open clusters with Hipparcos
(Robichon et al. \cite{Robichon et al.}) have substantially increased our
knowledge of the kinematic behaviour of the young stellar system in the
solar neighbourhood.

In Sect. \ref{samples} we describe the samples we have used, based on the
astrometric Hipparcos data plus a careful compilation of available radial
velocities and Str\"omgren photometry. This information enables us to
obtain reliable space velocities and individual ages for a large number of
stars. In Sect. \ref{Gould Belt} a maximum likelihood method is applied in
deriving the structural parameters and age of the Gould Belt. The velocity
field was studied by means of the classical first-order approach in Sect.
\ref{Oort}. The trends observed in the Oort constants are interpreted in
terms of the expansion of the Gould Belt and the influence of stellar
aggregates.

%
\section{The working sample}\label{samples}

Our initial sample (see Fern\'andez \cite{Fernandez} for more details)
contained 6922 O- and B-type stars ({\it Hipparcos Internal Proposal
INCA060} completed with all the O and B {\it survey} stars), of which 5846
belong to the Hipparcos {\it survey}. The observational data were taken
from the following sources:

\begin{itemize}

\item Astrometric data from the Hipparcos Catalogue (ESA \cite{ESA2}):  
positions in equatorial coordinates, parallaxes and proper motions,
together with their standard errors and the correlations between them.  
Given the mean standard error in the trigonometric parallax provided by
Hipparcos, reliable distances are available only up to 200-400 pc.

\item Str\"omgren photometry from Hauck \& Mermilliod's (\cite{Hauck et 
al.}) compilation for deriving individual photometric distances and ages.

\item Radial velocities from Grenier's (\cite{Grenier}) compilation plus
additional sources.

\end{itemize}

In the following subsections we describe the procedure used and the
accuracy achieved in the derivation of individual distances, spatial
velocities and ages, together with a discussion of the possible
observational biases present in the final working samples.

\subsection{Stellar distances}\label{distances}

To derive the best distance estimate for each star in the sample, an
analysis of individual errors, possible observational biases and
systematic differences between distances derived from Hipparcos
trigonometric parallaxes and photometric absolute magnitudes was
performed.

\begin{figure}
  \resizebox{\hsize}{!}{\includegraphics{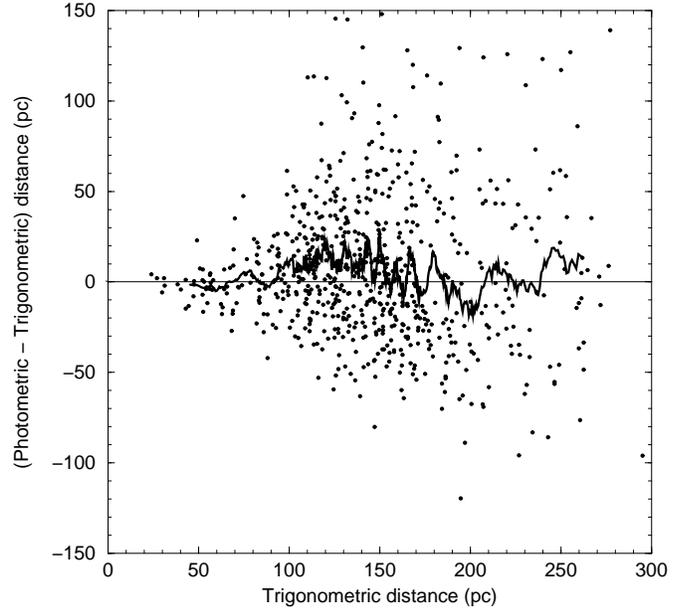}}
  \caption{Comparison of photometric and trigonometric distances for
           the sample stars with $\sigma_\pi/\pi < 0.15$. The solid line
           is a 25-point running average.}
  \label{fig.r.fot.trig}
\end{figure}

Only 3031 stars from our initial sample have complete Str\"omgren
photometry ($b-y$, $m_1$, $c_1$, $\beta$ and $V$) in Hauck \& Mermilliod's
(\cite{Hauck et al.}) catalogue. Recently, Kaltcheva \& Knude
(\cite{Kaltcheva et al.}) found that photometric distances derived from
Crawford's (\cite{Crawford}) and Balona \& Shobbrook's (\cite{Balona et
al.}) calibrations show good agreement with Hipparcos trigonometric
distances, and that no dependence on star rotation can be observed. The
comparison of distances derived using Crawford's calibration and Hipparcos
distances for the non-binary stars in our sample with $\sigma_\pi/\pi <
0.15$ (see Fig. \ref{fig.r.fot.trig}) shows no systematic trends. We also
verify that the use of another calibration (Balona \& Shobbrook
\cite{Balona et al.}; Jakobsen \cite{Jakobsen}) does not alter the
kinematic results presented in Sect. \ref{Oort}. After this analysis,
Crawford's (\cite{Crawford}) calibration was adopted to derive photometric
distances. A relative error in the photometric distance was computed
following Lindroos (\cite{Lindroos}). Depending on the spectral type and
luminosity class, this error ranges between 14-23\%.

Reliable photometric distances can only be derived for single,
non-variable and non-peculiar stars, so no photometric distance was
derived for those stars classified as double or multiple in the Hipparcos
Catalogue with a component separation $\rho < 10\arcsec$ and a magnitude
difference between components $\Delta H_{\mathrm{p}} < 3^{\mathrm{m}}$,
for variable stars with a variation in the Hipparcos magnitude system
$\Delta H_{\mathrm{p}} > 0.6^{\mathrm{m}}$ and for stars photometrically
classified as peculiar (Jordi et al. \cite{Jordi et al.}).

For those stars for which we could obtain trigonometric and photometric
distances, we used the distance with the smallest relative error. This
procedure was preferred to the derivation of an error-weighted mean of
both distance measurements since the latter would only systematically
reduce the error for those stars around 150-250 pc (as will be seen in
Sect. \ref{Oort}, this error is used as a weight in the condition
equations, so a different weight would be assigned as a function of
distance). Furthermore, negative trigonometric parallax or the bias in the
trigonometric distance discussed below -- non negligible for stars with
large relative error in parallax -- could not be controlled.

\begin{figure}
  \resizebox{\hsize}{!}{\includegraphics{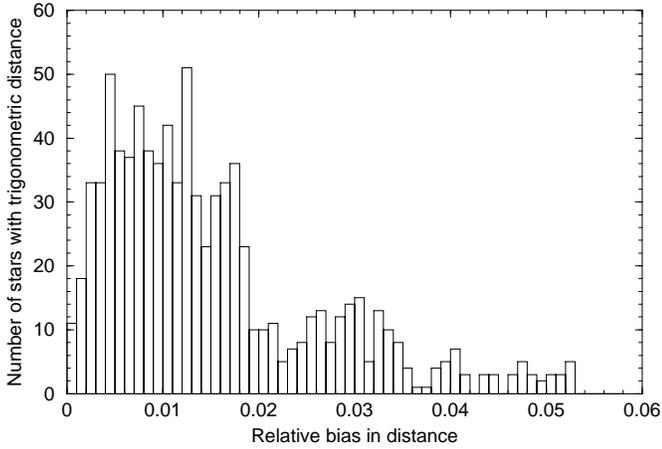}}
  \caption{Relative bias distribution for the 858 stars in our sample for
           which the trigonometric distance was chosen as providing the
           best distance determination.}
  \label{fig.bias.r.trig}
\end{figure}

In the case of stars for which the photometric distance was not available,
to avoid bias, the trigonometric distance was only accepted if its
relative error was smaller than 25\%. If distances are estimated as $R =
1/\pi$, a symmetric error law for parallaxes results in a non-symmetric,
biased distribution for distances. According to Arenou \& Luri
(\cite{Arenou et al.}), for small relative errors ($\la$ 25\%) and
assuming a Gaussian law for the error on the observed parallax, this bias
can be approximated by:

\begin{equation}\label{pi.bias}
  B(R) \approx \frac {1}{\pi_{\mathrm{t}}}
  \left( \frac {\sigma_{\pi}}{\pi_{\mathrm{t}}} \right)^2
\end{equation}

\noindent being $\pi_{\mathrm{t}}$ the {\it true} parallax. Individual
corrections for the observed parallaxes are not possible because the bias
is a function of the {\it true} parallax (Brown et al. \cite{Brown et
al.2}). However, an estimation of the effect of this bias in our
trigonometric distances may be made considering in Eq. (\ref{pi.bias})  
the {\it observed} instead of the {\it true} parallaxes. In Fig.  
\ref{fig.bias.r.trig} we show the relative bias distribution ($B(R)/R$)
for the 858 stars in our sample for which we chose the trigonometric
distance as the best distance estimate. This relative bias is always less
than 5.5\%, and is smaller than 3\% for 88\% of the stars, thus giving a
bias smaller than 5 pc for 82\% of the stars. Therefore, given the
impossibility of an individual correction of the biases, while recognizing
their smallness (comparing, for instance, with the relative errors in the
parallax), we may conclude that our assumption of considering directly the
trigonometric distance as given by $R = 1/\pi$ for stars with relative
error in parallax smaller than 25\% is a good approximation.

\subsection{Stellar radial velocities}\label{vrad}

Our main source of radial velocities was the compilation of Grenier
(\cite{Grenier}), who made a complete compendium and revision of the
compilations of Barbier-Brossat (\cite{Barbier-Brossat}) and Duflot et al.
(\cite{Duflot et al.}). Priority was given to Barbier-Brossat, and only
stars with A, B or C quality in Duflot et al. were considered. Using these
sources, 3397 stars from our initial sample have radial velocity
measurements. We rejected those stars with an individual error in the
radial velocity higher than 10 km \mbox{s$^{-1}$} (131 stars, i.e. 3.9\%
of the stars with radial velocity).

\begin{figure}
  \resizebox{\hsize}{!}{\includegraphics{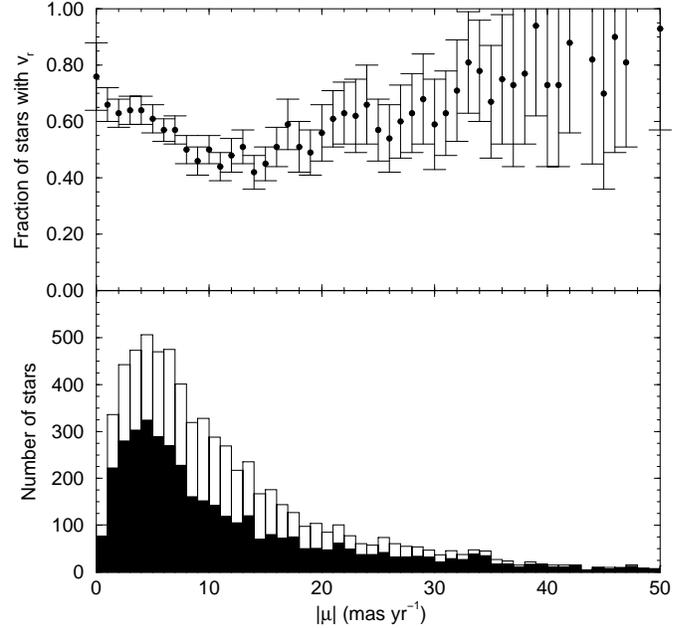}}
  \caption{Fraction of stars with radial velocity (top) and distribution
           of the stars with proper motion (blank histogram) and radial
           velocity (filled histogram) as a function of proper motion
           (bottom). Error bars were estimated from a Poissonian error
           distribution.}
  \label{fig.bias.vr.mu}
\end{figure}

\begin{figure}
  \resizebox{8.5cm}{!}{\includegraphics{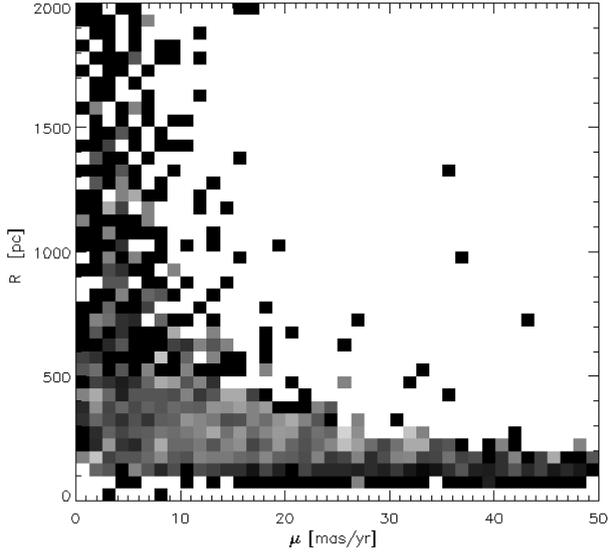}}
  \caption{Fraction of stars with radial velocity against proper motion
           and distance. This fraction is showed in a grey scale, from 0
           (white) to 1 (black).}
  \label{fig.grid.fracVr.pm.R}
\end{figure}

\begin{figure}
  \resizebox{8.5cm}{!}{\includegraphics{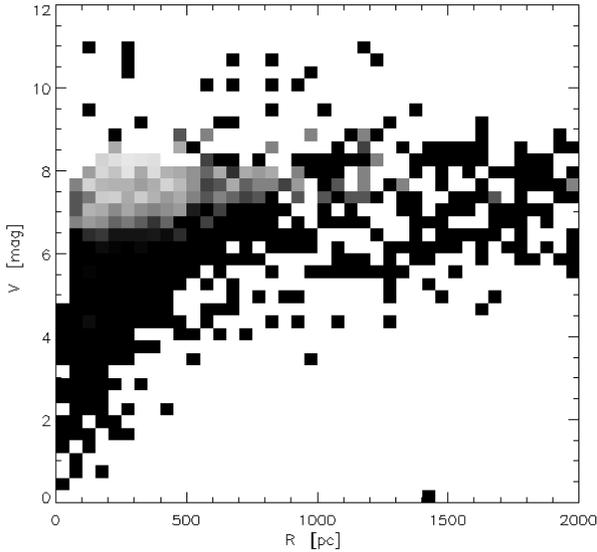}}
  \caption{Fraction of stars with radial velocity against distance and
           apparent visual magnitude. This fraction is showed in a grey
           scale, from 0 (white) to 1 (black).}
  \label{fig.grid.fracVr.R.V}
\end{figure}

Binney et al. (\cite{Binney et al.}), when working with nearby stars from
all spectral types, emphasized that, due to observational programmes,
radial velocity availability is higher for high-proper motion stars. In
our case, specific observational programmes were undertaken in parallel
with the Hipparcos mission to obtain radial velocity data for {\it survey}
early type Hipparcos stars. To evaluate the effects of the observational
constraints on our kinematical study, we plotted in Fig.
\ref{fig.bias.vr.mu} the fraction of stars with known radial velocities
($q_{V_{\mathrm{r}}}$) against the total proper motion.  We can see as
this fraction is not a flat function of $|\mu|$: it decreases for $|\mu|
\la$ 10 mas yr$^{-1}$ and rises for $|\mu| \ga$ 10 mas yr$^{-1}$. To
understand this effect in Fig. \ref{fig.grid.fracVr.pm.R}
$q_{V_{\mathrm{r}}}$ is shown against proper motion and distance. We can
see a higher degree of completeness for distant stars. From Fig.
\ref{fig.grid.fracVr.R.V} we derived a completeness limit of $V \approx
6.5$ for radial velocities, although nearly all far and faint stars ($R
\ga$ 1000 pc, so low $|\mu|$) have radial velocity data. A possible origin
for this effect is due to observing programmes devoted to open clusters
and associations. From this analysis we conclude that a kinematical bias
is present in our sample and, although its effects on the analysis
performed in Sect. \ref{Oort} are expected to be negligible, they must be
evaluated -- see \ref{appendix2} -- through numerical simulations.

\subsection{Stellar ages}

Individual ages were computed from the evolutionary models of Bressan et
al. (\cite{Bressan et al.}) for solar composition following the
interpolation algorithm described in Asiain et al. (\cite{Asiain et
al.1}). The algorithm considers, as input parameters, the
$T_{\mathrm{eff}}$ and $\log g$ derived from the Str\"omgren photometric
indices (Moon et al. \cite{Moon et al.}; Napiwotzki et al.
\cite{Napiwotzki et al.}).

The weakness in this procedure lies in the inability to take into account
the effects of stellar rotation when deriving ages from photometry.
Figueras \& Blasi (\cite {Figueras et al.2}), analysing a sample of main
sequence B7-A4 stars, found that actual photometric ages increment by
30-50\% in average, if rotation is not considered. As individual
corrections are not possible, and an important fraction of the O-B9
main-sequence stars are high rotators, this important systematic trend has
to be considered when deriving an age for the Gould Belt system.

\begin{figure}
  \resizebox{6.5cm}{!}{\includegraphics{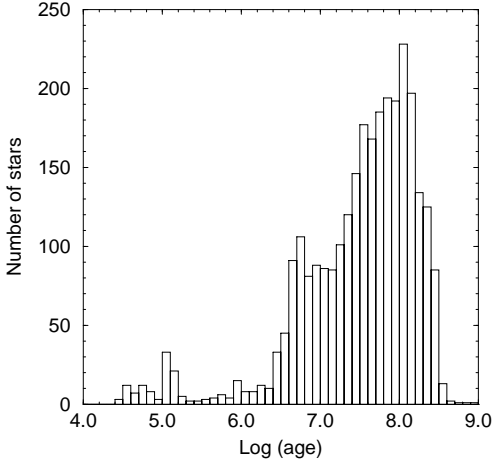}}
  \caption{Age distribution of the sample of O and B stars (2864 stars).}
  \label{fig.edat.hist}
\end{figure}

\begin{figure}
  \resizebox{6.5cm}{!}{\includegraphics{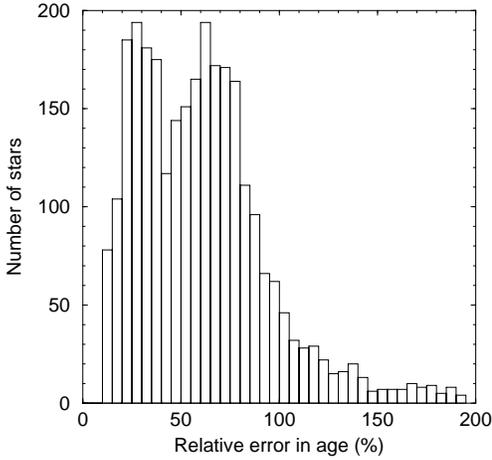}}
  \caption{Distribution of the relative error in age of the O and B stars
           in the sample. 88\% of the stars have a relative error in
           age smaller than 100\%.}
  \label{fig.error.edat.hist}
\end{figure}

First, individual ages were derived for all the stars for which
photometric data was available, without taking into account binarity,
variability or photometric peculiarities. The age and relative error in
age distributions computed for 2864 stars are presented in Fig.
\ref{fig.edat.hist} and Fig. \ref{fig.error.edat.hist}, respectively. With
the aim of retaining as many as possible of the very young stars in our
final sample, a careful treatment was followed to take into account the
effects of binarity, duplicity or peculiarity in the age computation:

\begin{itemize}

\item In the case of double or multiple stars for which only joint
photometry is available (i.e., systems with $\rho < 10\arcsec$ and $\Delta
H_{\mathrm{p}} < 3^{\mathrm{m}}$), the computed age is greater than the
real age (Trimble \& Ostriker \cite{Trimble et al.}). All double or
multiple systems with computed ages smaller than 30 Myr have an actual age
inside the interval $ \tau \le$ 30 Myr, so they were retained in the
working sample only if they have a reliable Hipparcos trigonometric
parallax ($\sigma_\pi/\pi < 0.25$), as the effects of duplicity render the
photometric distance meaningless. Furthermore, roughly 80\% of these stars
have $\tau + \sigma_{\tau} \le$ 60 Myr, so there is a high probability
that they belong to the Gould Belt. Double or multiple systems with an
estimated age larger than 30 Myr were rejected.

\item We rejected those stars with photometric or spectral peculiarities.
The photometric indices accounting for temperature for Bp stars are bluer
due to peculiarities, thus producing estimated ages smaller than real
(Hauck \cite{Hauck}). Variable stars with $\Delta H_{\mathrm{p}} >
0.6^{\mathrm{m}}$ were also rejected.

\item As shown in Fig. \ref{fig.error.edat.hist}, about 12\% of the sample
has a relative error in age larger than 100\%. From them, those placed
below the ZAMS (206 stars) are peculiar stars that are already rejected or
very young stars. We checked that the stars above the ZAMS had a computed
age smaller than about 30 Myr (116 stars). In any case, as these stars are
expected to be very young and therefore of great importance to our study
of the Gould Belt (they represent roughly 25\% of the stars with $\tau \le
$ 30 Myr), we decided to retain them in the final sample.

\end{itemize}

\subsection{Working samples}

\begin{figure}
  \resizebox{\hsize}{!}{\includegraphics{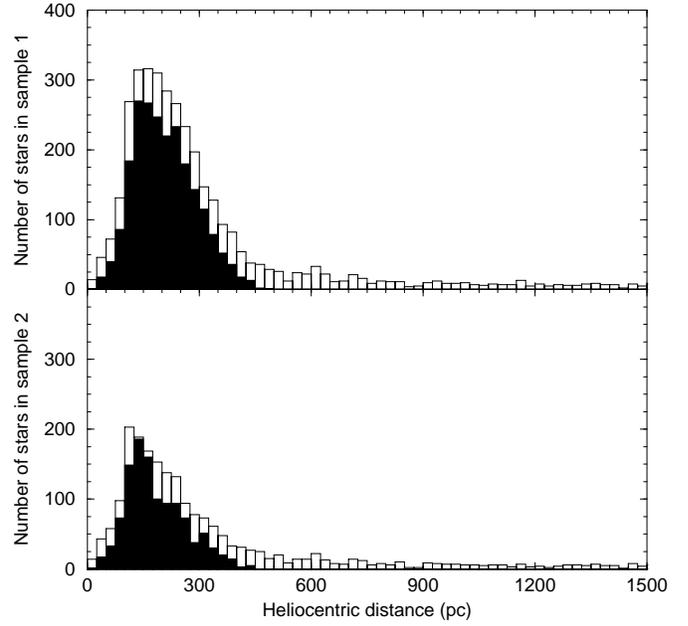}}
  \caption{Distance distribution of the samples 1 (top) and 2 (bottom) of
           O and B stars defined in the text. Filled histograms show the
           stars for which Hipparcos trigonometric distance was used
           (56.3\% and 50.2\% of the stars belonging to samples 1 and 2,
           respectively).}
  \label{fig.r.hist}
\end{figure}

\begin{table}[top]
   \caption{Averaged errors in the samples of O and B stars.}
   \label{tab.error.mostres}
\begin{tabular}{cll}
\hline
{\bf Error}                                         & {\bf Sample 1}      & {\bf Sample 2}     \\
\hline
\hline
$\overline{\sigma}_{\pi}$                           & 0.60 mas            & 0.57 mas           \\
$\overline{\left( \frac {\sigma_\pi}{\pi} \right)}$ & 0.168               & 0.163              \\
$\overline{\sigma}_{\mu_{\alpha} \cos \delta}$      & 0.83 mas yr$^{-1}$  & 0.81 mas yr$^{-1}$ \\
$\overline{\sigma}_{\mu_{\delta}}$                  & 0.70 mas yr$^{-1}$  & 0.67 mas yr$^{-1}$ \\
$\overline{\sigma}_{v_{r}}$                         & ---                 & 3.44 km s$^{-1}$   \\
\hline
\end{tabular}
\end{table}

\begin{figure}
  \resizebox{\hsize}{!}{\includegraphics{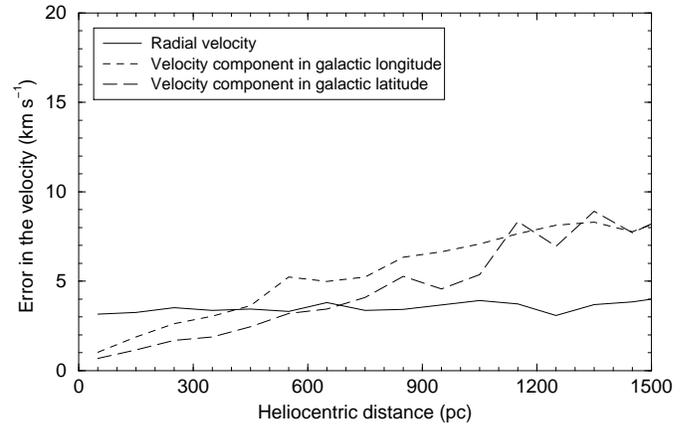}}
  \caption{Averaged errors in the three velocity components plotted 
           against heliocentric distance.}
  \label{fig.error.v.mmobil}
\end{figure}

Following the procedure described above, two samples of stars were formed:

\begin{itemize}

\item {\bf Sample 1:} Containing 3915 stars with known distance and 
proper motions.

\item {\bf Sample 2:} A subsample of sample 1 containing 2272 stars with 
known distance, radial velocity and proper motions.

\end{itemize}

\noindent In Fig. \ref{fig.r.hist} we show their distance distribution.
Although the initial sample contains all the {\it survey} Hipparcos stars
(complete up to $V$ = 7.9), the lack of photometry and radial velocity
data reduces the completeness of the samples to about $ V_{\mathrm{lim}}
\approx$ 6.3, that is up to 150 pc for a B9V star, the faintest in our
sample.

The averaged errors of several quantities are presented in Table
\ref{tab.error.mostres}. In Fig. \ref{fig.error.v.mmobil} we show the
averaged errors in the three velocity components plotted against
heliocentric distance (computed taking into account the correlations
between the different variables provided by the Hipparcos Catalogue).

\begin{figure*}
  \resizebox{\hsize}{!}{\includegraphics{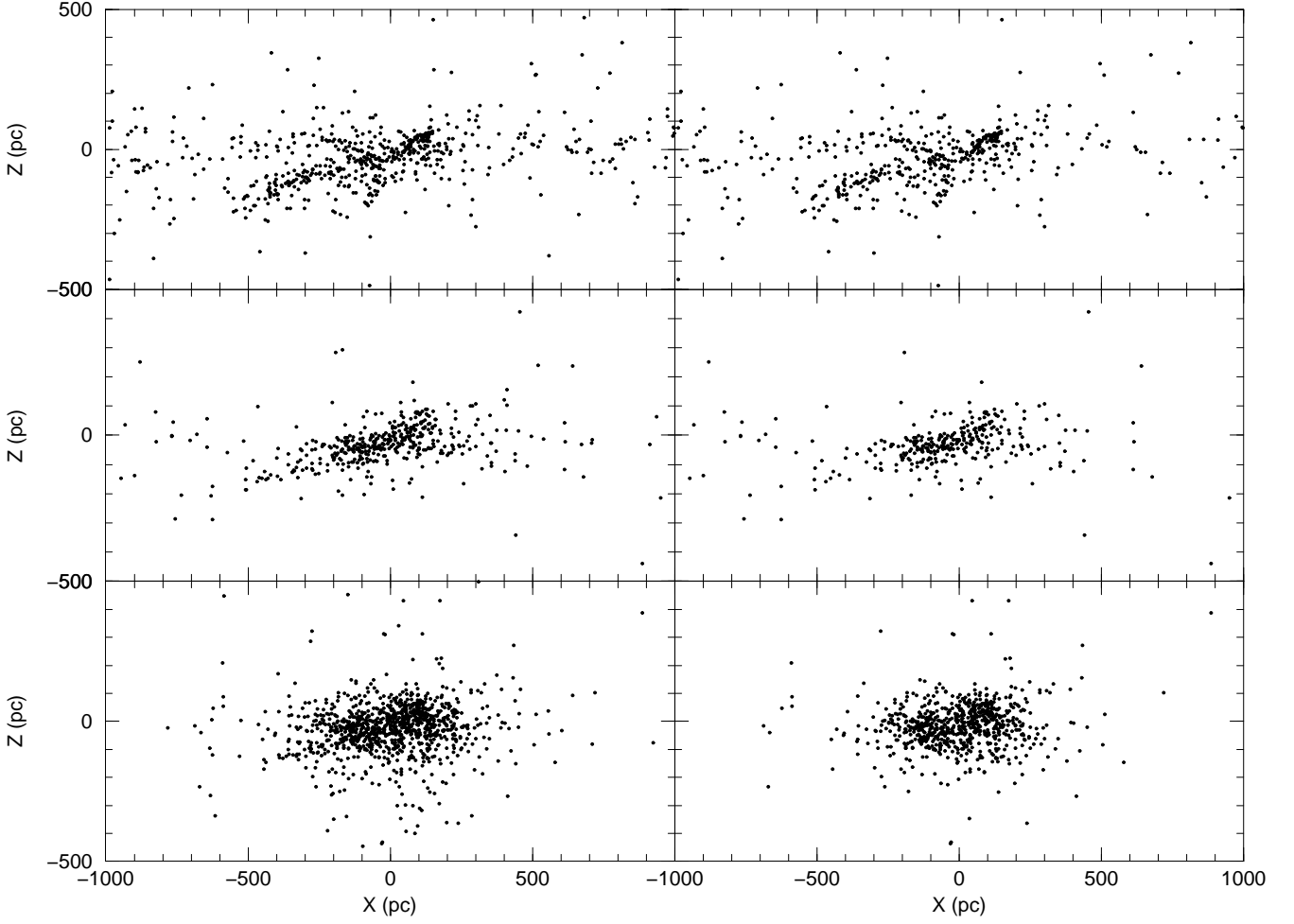}}
  \caption{Distribution on the $X$-$Z$ galactic plane for the stars in the
           samples 1 (left) and 2 (right). At the top, the stars with an
           age less than 30 Myr; in the middle, those with an age between
           30 Myr and 60 Myr; and at the bottom, those with an age larger
           than 60 Myr.}
  \label{fig.xz}
\end{figure*}

The number of stars in samples 1 and 2 is reduced to 2468 and 1789,
respectively, when individual ages are required. In Fig. \ref{fig.xz} we
show the position of these stars projected on the $X$-$Z$ plane ($X$
positive towards the galactic center and $Z$ towards the north galactic
pole), classified in three age groups ($\tau \le$ 30 Myr, 30 $< \tau \le$
60 Myr and $\tau >$ 60 Myr). The Gould Belt is recognized as a tilted
structure with regard to the galactic plane ($Z = 0$), mainly in the
region with $X < 0$ and $Z < 0$. This structure is clearly visible for
stars younger than 60 Myr. The presence of some stars belonging to this
structure in the interval 60 $< \tau \le$ 90 Myr is fully justified by the
large errors in the estimate of individual ages.

%
\section{The Gould Belt}\label{Gould Belt}

\subsection{Structural parameters of the Gould Belt}

\begin{table*}[top]
   \caption{Structural parameters of the Gould Belt as a function of
            distance and age: inclination ($i_{\mathrm{G}}$), longitude
            of the ascending node ($\Omega_{\mathrm{G}}$), fraction of
            stars belonging to the Gould Belt ($q$), angular halfwidth of
            the Gould Belt ($\xi_{\mathrm{G}}$) and the galactic belt
            ($\xi_{\mathrm{g}}$), and number of stars ($N$). Limiting
            visual apparent magnitude: 7.0.}
   \label{tab.par.Gould}
\begin{tabular}{ccccccr}
 \hline
$R$ (pc) & $i_{\mathrm{G}}$ ($\degr$) & $\Omega_{\mathrm{G}}$ ($\degr$) & $q$ & $\xi_{\mathrm{G}}$ ($\degr$) & $\xi_{\mathrm{g}}$ ($\degr$) & $N$ \\
\hline
\hline
\multicolumn{7}{c}{$\tau \le$ 30 Myr} \\
\hline
\hline
\mbox{$R \le$ 400}           & 21.2$_{(1.3)}$  & 287.3$_{(4.2)}$   & 0.60 &  6.2 & 22.5 & 236 \\
\mbox{$R \le$ 600}           & 19.9$_{(1.6)}$  & 282.8$_{(5.2)}$   & 0.66 &  7.0 & 22.6 & 300 \\
\mbox{600 $< R \le$ 2000}    & 11.8$_{(2.2)}$  & 316.1$_{(10.7)}$  &      &      &      & 126 \\
\hline
\hline
\multicolumn{7}{c}{30 $< \tau \le$ 60 Myr} \\
\hline
\hline
\mbox{$R \le$ 400}           & 15.9$_{(2.5)}$  & 294.9$_{(6.5)}$   & 0.64 &  7.3 & 23.7 & 261 \\
\mbox{$R \le$ 600}           & 15.5$_{(2.6)}$  & 293.4$_{(6.5)}$   & 0.62 &  7.2 & 22.5 & 297 \\
\mbox{600 $< R \le$ 2000}    & 11.9$_{(22.7)}$ & 192.6$_{(164.1)}$ &      &      &      &  31 \\
\hline
\hline
\multicolumn{7}{c}{60 $< \tau \le$ 90 Myr} \\
\hline
\hline
\mbox{$R \le$ 400}           & 22.3$_{(2.1)}$  & 276.1$_{(4.9)}$   & 0.44 &  7.6 & 25.8 & 177 \\
\mbox{$R \le$ 600}           & 22.1$_{(4.1)}$  & 276.7$_{(4.1)}$   & 0.42 &  7.1 & 25.7 & 198 \\
\hline
\hline
\multicolumn{7}{c}{90 $< \tau \le$ 120 Myr} \\
\hline
\hline
\mbox{$R \le$ 400}           &  5.0$_{(32.8)}$ & 316.7$_{(378.4)}$ &      &      &      & 160 \\
\mbox{$R \le$ 600}           &  3.4$_{(33.8)}$ & 319.3$_{(569.2)}$ &      &      &      & 170 \\
\hline
\end{tabular}
\end{table*}

A classic problem in the study of the Gould Belt has been how to separate
the stars belonging to this structure from those belonging to the galactic
belt. Stothers \& Frogel (\cite{Stothers et al.}) and Taylor et al.
(\cite{Taylor et al.}) proposed different algorithms based on the
individual assignation of the stars to either belt. More recently,
Cabrera-Ca\~no et al. (\cite{Cabrera-Cano et al.}; hereafter CEA) proposed
{\it least mean classification error} decision criteria to separate those
stars belonging to each belt. These criteria, based on the spatial
distribution of the stars, run into an obstacle: the classification of
stars lying in the overlap region between both belts.

In this paper, we followed an alternative approach proposed by CTG. The
method assumes that the belts form two great circles in the celestial
sphere, with a star density which decreases with the angular distance from
each equator belt. The decrease in star density is assumed to follow a
Gaussian law, the standard deviation being the angular halfwidth of the
belt. Therefore, we suppose that the density distribution of the sample in
the celestial sphere can be written as:
\begin{equation}
  \sigma (l,b) = \sigma_{\mathrm{G}} (l,b) + \sigma_{\mathrm{g}} (l,b)
\end{equation}
where $\sigma_{\mathrm{G}}$ and $\sigma_{\mathrm{g}} $ are the density
distributions around the Gould Belt and the galactic belt equators,
respectively.

This decomposition allows us to derive several parameters for the Gould
Belt: the spatial orientation with regard to the galactic plane
($i_{\mathrm{G}}$, $\Omega_{\mathrm{G}}$), the fraction of stars belonging
to the Gould Belt ($q$), and the angular halfwidth of each belt
($\xi_{\mathrm{G}}$, $\xi_{\mathrm{g}}$). Comer\'on (\cite{Comeron1}),
using pre-Hipparcos data, showed that the $q$ parameter describes in an
appropriate way the main characteristics of the distribution of young
stars, in particular the extent of the belts as a function of galactic
longitude.

The resolution procedure, based on the maximum likelihood method, can be
found in Comer\'on (\cite{Comeron1}). Additionaly, an iterative procedure
until convergence was implemented here to minimize the dependence of the
final results on the departing values. As explained in Comer\'on
(\cite{Comeron1}), the method requires the homogenous completeness of the
sample throughout over the entire celestial sphere. Unfortunatelly,
although Hipparcos is complete up to $V =$ 7.9, the absence of photometric
measurements reduces this limit, so substantially reducing the distance
limit of the intrinsically faint stars. As a good compromise between the
constraints of completeness and the need for a statistically
representative number of stars we considered only those stars that are
brighter than $V =$ 7.0 in sample 1. Numerical simulations, presented in
\ref{appendix1}, allowed us both to assess how these incompleteness
effects could influence the results and to have an external evaluation of
the errors on the derived structural parameters.
  
The results obtained for the real sample are presented in Table
\ref{tab.par.Gould}. We see that the Gould Belt's structure is clearly
detected in the subsamples of young stars with $R \le$ 600 pc. From the
top left panel in Fig. \ref{fig.xz} (stars with $\tau \le$ 30 Myr) we also
see that this structure extends up to 600 pc into the south galactic
hemisphere and only up to 200-300 pc in the north. As young stars in the
galactic plane reach distances greater than 1000 pc without having any
substantial decrease in density, we can state that the distance cut off
derived for this structure is real in this age interval and not a
consequence of the incompleteness of our sample. An extent of about 600 pc
is in agreement with Lindblad et al. (\cite{Lindblad et al.2}), who
assumed that the prominent associations in the Gould Belt are within a
distance of 700 pc.

The orientation parameters were found to be $i_{\mathrm{G}} =
16$-$22\degr$ and $\Omega_{\mathrm{G}} = 275$-$295\degr$. The orientation
parameters are maintained up to the interval 60-90 Myr. For stars older
than 90 Myr the method does not converge -- large errors are found for
$i_{\mathrm{G}}$ and $\Omega_{\mathrm{G}}$, and $q$ and halfwidths are
undetermined --, so we conclude that the Gould Belt is no longer present.
However, at this stage, it is mandatory to verify that the dissapearance
of the structure for older stars is not a consequence of our observational
constraints (in our sample, stars with $\tau \approx$ 60 Myr have a
limited distance about 400 pc). Simulations show that distance cut off for
these stars does not significantly disturb the determination of structural
parameters. In other words, if a substantial number of stars with ages
larger than 60 Myr would be present in the Gould Belt, our algorithm would
be able to detect them from the number of stars available at present. The
smaller $i_{\mathrm{G}}$ value derived in the interval 30-60 Myr compared
with that obtained in the intervals $\tau \le$ 30 Myr and 60 $< \tau \le$
90 Myr has no explanation at present (it cannot be attributed to the
sample distance horizon). From the simulations, an uncertainty of
3-$6\degr$ is expected for $i_{\mathrm{G}}$ in these age intervals. In any
case, our values for the orientation parameters are in good agreement with
those published in the literature. Lesh (\cite{Lesh}), Stothers \& Frogel
(\cite{Stothers et al.}) and Westin (\cite{Westin}) reported values of
$i_{\mathrm{G}} = 19$-$22\degr$ and $\Omega_{\mathrm{G}} =
270$-$300\degr$. CTG reported $i_{\mathrm{G}} = 22.3\degr$ and
$\Omega_{\mathrm{G}} = 284.5\degr$ from a sample of O-A0 stars. Recently,
Guillout et al. (\cite{Guillout et al.1}) analysed the sky distribution of
X-ray emitting stars belonging to the RASS-Tycho sample and found a low
galactic latitude feature -- the Gould Belt -- with an orientation with
regard to the galactic plane of $i_{\mathrm{G}} = 27.5 \pm 1\degr$ and
$\Omega_{\mathrm{G}} = 282 \pm 3\degr$. Even if we take the errors into
account, this inclination value is not compatible with ours. A possible
explanation for this discrepancy is the fact that Guillout et al.'s sample
was restricted to stars that were not located at such great distances ($R
<$ 200 pc), where the most prominent structure belonging to the Gould Belt
-- the Sco-Cen complex -- defines a slightly higher slope than those
located at a greater distance and in the opposite direction (mainly the
Ori OB1 association). CEA, after identifying those stars belonging to the
Gould Belt, found $i_{\mathrm{G}} = 17.5$-$18.3\degr$ and
$\Omega_{\mathrm{G}} = 287$-$294\degr$, which is in very close agreement
with our results.

For stars younger than 60 Myr, the fraction of stars belonging to the
Gould Belt ($q$) were found to be 0.60-0.66. This value of $q$ decreased
to 0.42-0.44 when considering stars with an age between 60 and 90 Myr.
From \ref{appendix1} we expect an uncertainty of $\approx0.15$ in $q$.
Therefore, its decrease for old stars must be real. CEA, classifying by
spectral type groups, found $q =$ 0.44 for O-B2.5 stars with $R <$ 1000
pc, and $q =$ 0.36 for O-B9.5 stars.

The angular halfwidths of the belts were found to be $\xi_{\mathrm{G}} =
6$-$8\degr$ and $\xi_{\mathrm{g}} = 22$-$26\degr$ (see Table
\ref{tab.par.Gould}) with an approximated uncertainty of $5\degr$ (see
\ref{appendix1}). These values tend to increase smoothly when we go from
young to not so young stars. This effect, if real, could be simply due to
the geometrical effect produced by the decrease in the mean distance when
older stars are considered. We found that the Gould Belt is narrower than
the galactic belt, contrary to the result obtained by CTG, and in perfect
agreement with the relationship 1:3 obtained by Stothers \& Frogel
(\cite{Stothers et al.}) studying the scale heights for both belts. The
same trend is observed in the paper by CEA.

\subsection{The age of the Gould Belt}\label{sect.age}

On the basis of the spatial distribution analysed in the previous section,
and using individual photometric ages, we estimate the belt to be younger
than 60 Myr. As explained before there are two major biases in the
computation of individual photometric ages that account for the presence
of certain stars belonging to the Gould Belt in the age interval 60-90
Myr. First, due to the significant uncertainties in the age computation,
we expect that some stars with actual ages smaller than 60 Myr are
included in this interval. Second, and more importantly, the fact that we
cannot include the effects of stellar rotation on the age computation
scheme means there is a systematic increase that can be evaluated in 30-40
Myr (Figueras \& Blasi \cite{Figueras et al.2}).

The age estimates for the Gould Belt found in the literature lie in the
interval 20-90 Myr. Lesh (\cite{Lesh}) found an age of 45 Myr when
supposing two superposed stellar distributions, and 90 Myr when
considering only one homogeneous population in expansion. Lindblad et al.
(\cite{Lindblad et al.1}) suggested an age of 30-40 Myr, the system being
born in a spiral arm. Franco et al. (\cite{Franco et al.}) found an age of
60 Myr for the Orion and Monoceros molecular cloud complexes. Tsioumis \&
Fricke (\cite{Tsioumis et al.}) and Comer\'on \& Torra (\cite{Comeron et
al.1}) reported 60 Myr and 70 Myr, respectively, from kinematic studies.
From stellar individual age determinations using Str\"omgren photometry,
Westin (\cite{Westin}) found an upper limit of 60 Myr. CTG divided their
sample of O and B stars in spectral type subsamples, and reported a lower
limit for age equal to the lifetime of a B4-type star (about 50 Myr).
Guillout et al. (\cite{Guillout et al.2}), analysing the RASS-Tycho
sample, reported the detection of a very young active late-type stellar
population belonging to the Gould Belt. The X-ray luminosity distribution
was compatible with an age of 30-80 Myr for these stars. More recently,
Moreno et al. (\cite{Moreno et al.}) derived an age of 20 Myr, assuming
that the Gould Belt was formed by an expanding shell.

A parallel determination of the Gould Belt's age from its kinematics will
be given in Sect. \ref{sect.gould}.

%
\section{Kinematics of young stars in the solar neighbourhood}\label{Oort}

\subsection{Fit of the kinematic model}

\subsubsection{Equations for the linear model}

The Oort constants were derived using the first-order development of the
systematic velocity field:
\begin{eqnarray}
  V_{\mathrm{r}} & = & A \, R \, \sin{2l} \cos^2{b}
\nonumber \\
      & & + C \, R \, \cos{2l} \cos^2{b} + K \, R \, \cos^2{b}
\nonumber \\
      & & - U_\odot \cos{l} \cos{b} - V_\odot \sin{l} \cos{b} - W_\odot \sin{b}
\label{eq.vr}
\nonumber \\
\\
  R \, k \, \mu_{\mathrm{l}} \cos{b} & = & A \, R \, \cos{2l} \cos{b}
      + B \, R \, \cos{b}
\nonumber \\
      & & - C \, R \, \sin{2l} \cos{b}
\nonumber \\
      & & + U_\odot \sin{l} - V_\odot \cos{l}
\label{eq.vl}
\\
\nonumber \\
  R \, k \, \mu_{\mathrm{b}} & = & - A \, R \, \sin{2l} \sin{b} \cos{b}
\nonumber \\
      & & - C \, R \, \cos{2l} \sin{b} \cos{b} - K \, R \, \sin{b} \cos{b}
\nonumber \\
      & & + U_\odot \cos{l} \sin{b} + V_\odot \sin{l} \sin{b} - W_\odot \cos{b}
\label{eq.vb}
\nonumber \\
\end{eqnarray}
\noindent where $l$, $b$ are the galactic coordinates, $R$ the
heliocentric distance in pc, $V_{\mathrm{r}}$ the radial velocity in km
s$^{-1}$ and $\mu_{\mathrm{l}}$, $\mu_{\mathrm{b}}$ the proper motions in
$\arcsec$ yr$^{-1}$ of each star. The constant $k$ = 4.741 km yr (s pc
$\arcsec$)$^{-1}$. $U_\odot$, $V_\odot$ and $W_\odot$ are the components
of the peculiar motion of the Sun in km s$^{-1}$ with regard to the
circular velocity and $A$, $B$, $C$ and $K$ are the Oort constants, linear
combinations of the gradients of the systematic velocity. No systematic
motions perpendicular to the galactic plane were considered other than
that arising from the solar peculiar motion. As recently verified by
Palou\u{s} (\cite{Palous}), the $E$, $D$ and $H$ terms describing this
systematic motion do not improve the results and their values are of low
significance.

\subsubsection{Resolution procedure}

A weighted least squares fit was performed to estimate the model
parameters from Eqs. (\ref{eq.vr}),(\ref{eq.vl}) and (\ref{eq.vb}), taking
the residual velocity of the star as a random error. Consequently, the
weight of each equation was chosen as (Cr\'ez\'e \cite{Creze2}):

\begin{equation}
  p_i = \frac {1} {\sigma_{i,\mathrm{obs}}^2 + \sigma_{i,\mathrm{cos}}^2}
\end{equation}
 
\noindent where $\sigma_{\mathrm{obs}}$ are the individual observational
errors in each velocity component of the star, calculated by taking into
account the correlations between the different variables provided by
Hipparcos Catalogue, and $ \sigma_{\mathrm{cos}}$ is the projection of the
cosmic velocity dispersion ellipsoid ($ \sigma_{\mathrm{U}}$, $
\sigma_{\mathrm{V}}$, $ \sigma_{\mathrm{W}}$) in the direction of the
velocity component considered.  The detailed iterative procedure applied
to simultaneously derive the model parameters and the cosmic dispersion
for each of the subsamples considered is explained in Sect. \ref{cosmic}.

To check the quality of the least-squares fits we considered the $\chi^2$
statistics for $N - M$ degrees of freeedom, defined as:

\begin{equation}
  \chi^2 = \sum_{i=1}^{N} 
  \frac {\left[ y_i - y(x_i; a_1,... a_M) \right]^2}
  {\sigma_{i,\mathrm{obs}}^2 + \sigma_{i,\mathrm{obs}}^2}
\end{equation}

\noindent where $x_i$ are the independent data (sky coordinates and
distances), $y_i$ the dependent data (radial and tangential velocity
components), $N$ the number of equations and $M$ number of parameters to
be fitted. According to Press et al. (\cite{Press et al.}), if the
uncertainties (cosmic dispersion and observational errors) are
well-estimated, the value of $\chi^2$ for a moderately good fit would be
$\chi^2 \approx N - M$, with an uncertainty of $\sqrt{2 \; (N - M)}$.

To eliminate the possible outliers present in the sample due to both the
existence of high residual velocity stars (Royer \cite {Royer}) or stars
with unknown large observational errors, we rejected those equations with
a residual larger than 3 times the root mean square residual of the fit
(computed as $ \sqrt{[ y_i - y(x_i; a_1,... a_M)]^2 / N}$) and
recomputed a new set of parameters. We checked than no more than 10-15
stars from the total sample were rejected using this procedure, so
ensuring the objectivity of the rejection criterion.

Three distance intervals were considered: 100 $< R \le$ 600 pc, 600 $< R
\le$ 2000 pc and 100 $< R \le$ 2000 pc. The first two intervals allow us
to determine the influence of the Gould Belt on the stellar kinematics,
since as we have seen there is no evidence of the presence of this
structure for $R >$ 600 pc. The last interval provides a global vision of
the kinematics of young stars in the solar neighbourhood. Stars with $R
\le$ 100 pc were not considered because, on the one hand, they do not give
us information about galactic rotation and, on the other hand, peculiar
motions of these nearby stars (small errors, so large weight) can disturb
the results.

\begin{table*}[top]
   \caption{Oort constants and  solar motion for stars in the
            sample 2 with 600 $< R \le$ 2000 pc. Units: $A$, $B$, $C$, $K$
            in km s$^{-1}$ kpc$^{-1}$; $U_\odot$, $V_\odot$, $W_\odot$,
            $\sigma$ in km s$^{-1}$. $\chi^2/N_{\mathrm{eq}}$ is the value
            of $\chi^2$ divided by the number of equations. $N$ is the
            number of stars. The errors in the fitted parameters were 
            computed as $\sigma_i^2 = \sigma^2 C_{ii}^{-1}$, where 
            $C_{ii}^{-1} = Cov_{ii} / \sigma^2$ is the corresponding
            element in the covariance matrix and $\sigma$ the standard
            deviation of the measurements with unit weight (Linnik
            \cite{Linnik}). As expected, large errors in $K$ and $W_\odot$
            are obtained when derived from proper motion data or radial
            velocity alone, respectively. A constant cosmic dispersion of
            ($\sigma_U$, $\sigma_V$, $\sigma_W$) = $(8,8,5)$ km s$^{-1}$
            was considered in all cases.}
   \label{tab.constants.vr.mp.vr+mp.Oort}
\begin{tabular}{crrrrrrrccc}
\hline
Components          &        $A$      &        $B$        &        $C$       & $K$
                    &     $U_\odot$   &     $V_\odot$     &     $W_\odot$    & $\sigma$ & $\chi^2/N_{\mathrm{eq}}$ & $N$      \\
\hline
\hline
& \multicolumn{10}{c}{600 $< R \le$ 2000 pc} \\
\hline
\hline
$V_{\mathrm{r}}$    &  11.7$_{(1.1)}$ &                   & $-$0.7$_{(1.1)}$ & $-$3.0$_{(0.7)}$
                    &   7.2$_{(1.3)}$ &    16.6$_{(1.2)}$ &    7.4$_{(3.2)}$ & 14.3    &  2.64    &  308 \\
$\mu_{\mathrm{l}} +
 \mu_{\mathrm{b}}$  &  14.5$_{(0.9)}$ & $-$12.8$_{(0.6)}$ &    0.7$_{(0.9)}$ & $-$0.7$_{(3.2)}$
                    &  10.7$_{(1.0)}$ &    12.1$_{(0.9)}$ &    8.5$_{(0.5)}$ & 11.9    &  2.05    &  308 \\
$V_{\mathrm{r}}$ + $\mu_{\mathrm{l}} + 
 \mu_{\mathrm{b}}$  &  13.0$_{(0.7)}$ & $-$12.7$_{(0.8)}$ &    0.0$_{(0.7)}$ & $-$3.3$_{(0.7)}$
                    &   8.7$_{(0.9)}$ &    14.1$_{(0.8)}$ &    8.6$_{(0.6)}$ & 12.7    &  2.39    &  308 \\
\hline
\hline
& \multicolumn{10}{c}{600 $< R \le$ 2000 pc excluding the region with $200 < l < 250\degr$} \\
\hline
\hline
$V_{\mathrm{r}}$    &  14.0$_{(1.6)}$ &                   & $-$1.0$_{(1.2)}$ & $-$1.4$_{(1.0)}$
                    &  10.1$_{(1.7)}$ &    18.5$_{(1.6)}$ &    8.2$_{(4.0)}$ & 15.1    &  2.91    &  224 \\
$\mu_{\mathrm{l}} + 
 \mu_{\mathrm{b}}$  &  13.5$_{(1.1)}$ & $-$11.8$_{(0.9)}$ & $-$1.1$_{(1.4)}$ & $-$9.4$_{(4.2)}$
                    &   8.6$_{(1.4)}$ &    13.7$_{(1.3)}$ &    8.9$_{(0.6)}$ & 13.1    &  1.19    &  224 \\
$V_{\mathrm{r}}$ +
$\mu_{\mathrm{l}} +
 \mu_{\mathrm{b}}$  &  13.5$_{(1.0)}$ & $-$11.4$_{(1.0)}$ & $-$1.1$_{(0.9)}$ & $-$2.3$_{(0.9)}$
                    &   8.5$_{(1.2)}$ &    15.8$_{(1.0)}$ &    8.9$_{(0.8)}$ & 13.5    &  1.25    &  224 \\
\hline
\end{tabular}
\end{table*}

At this stage it is appropiate to clarify the physical meaning of the $A$,
$B$, $C$ and $K$ terms in the two regions we want to analyse. In the
region 600 $ < R \le$ 2000 pc, not affected by the Gould Belt, these terms
are expected to reflect the local shape of the rotation curve. Thus,
assuming a smooth variation in space, they are properly called {\it Oort
constants} and they account for the divergence ($K$), vorticity ($B$) and
shear ($A$: azimuthally, $C$: radially) of the general velocity field of
the galactic disk in the solar neighbourhood. Olling \& Merrifield
(\cite{Olling et al.}) distinguished between {\it Oort constants} for the
local shape of the rotation curve and {\it Oort functions} when accounting
for their variation with the galactic radius.  On the contrary, in the
region 100 $ < R \le$ 600 pc, these first derivatives of the velocity
field will include the peculiar velocity field associated with the Gould
Belt. Thus, strictly speaking, they should not be called {\it Oort
constants} and so we will refer to them as {\it Oort parameters}.

Before analyzing the different fits performed there are two interrelated
aspects that deserve special attention: the possible biases in the fitting
parameters induced by the characteristics of the sample and our
observational constraints -- irregular spatial distribution of the stars,
incompleteness effects, biases in the availability of radial velocity data
(already discused in Sect.2.2), etc.  -- and the systematic errors in the
parameters induced by the presence of observational errors in the right
hand side of the Eqs. (\ref{eq.vr}), (\ref{eq.vl}) and (\ref{eq.vb}), not
considered in our least square fit. Cr\'ez\'e (\cite{Creze1}) derived
approximate analytical corrections to evaluate the second aspect but, as
he stated, we think that numerical experiments are the best way to
globally evaluate and discuss all the effects present in our real sample.
Furthermore, these simulations can give us new insights in a long standing
problem: the discrepancies appearing between the solutions obtained using
radial velocity and proper motion equations. Following that, we performed
several numerical simulations that are detailled in \ref{appendix2}. The
results are commented, in the next paragraphs, along with the results from
real data.

In Table \ref{tab.constants.vr.mp.vr+mp.Oort}, we compare the solutions
obtained when considering only radial velocity data -- Eq. (\ref{eq.vr})  
--, proper motion data -- Eqs. (\ref{eq.vl}) and (\ref{eq.vb}) solved
simultaneously -- or the combined solution -- Eqs. (\ref{eq.vr}),
(\ref{eq.vl}) and (\ref{eq.vb})  -- for those stars in sample 2 with 600
$< R \le$ 2000 pc. As can be seen, a difference about 2-3 km s$^{-1}$
kpc$^{-1}$ in the $A$ value is present between the radial velocity and the
proper motion solutions when solved separately. One effect that could
reduce this discrepancy was pointed out by Cr\'ez\'e (\cite{Creze1}): a
non-negligible standard error in distances produces an understimation of
the $A$ Oort constant when derived from radial velocities. Our numerical
experiments (see Table \ref{tab.sim}) indicate that this effect is less
important than the bias induced by the cut in the observed distance
(affected by errors), which produces a bias of about 1-1.5 km s$^{-1}$
kpc$^{-1}$ in the opposite sense. From these simulations we conclude that
the difference present in the real sample could even be enlarged if the
observational bias could be removed.

Recently, Feast et al. (\cite{Feast et al.2}), working with a sample of
Hipparcos Cepheids, showed that the discrepancy appearing between the $A$
value derived from proper motions (Feast \& Whitelock \cite{Feast et
al.1}) and that derived from radial velocities (Pont et al. \cite{Pont et
al.}) disappears when the new Hipparcos Cepheid distance scale is
considered in the last equations. In our case, we verified that an
overestimation in our photometric distances by a factor of 20\% -- e.g.
assuming a stellar rotation effect in the absolute magnitude derivation
(Lamers et al. \cite{Lamers et al.}, Domingo \& Figueras \cite{Domingo et
al.}) -- can account only for a difference of 1-2 km s$^{-1}$ kpc$^{-1}$
between both solutions.

Lindblad et al. (\cite{Lindblad et al.2}) attributed this discrepancy to
the irregular distribution of stars and stellar groups and also to a
possible non-linearity in the velocity field. As explained in
\ref{appendix2} our simulations take into account the irregular spatial
distribution of real stars and we saw there that the discrepancy did not
appear. More promissing is the hypothesis that the discrepancies come from
departures of some stellar groups from the adopted linear model. In our
sample, we verified that when the stars in the region $200 < l \le
250\degr$ (partly composed by stars of the Ori OB1 and Col 121
associations) are removed, the discrepancy on $A$ vanishes (see Table
\ref{tab.constants.vr.mp.vr+mp.Oort}). We also confirmed that when other
special regions are removed, the discrepancy is maintained.  Another
important point we realize when studying these results is the variation in
$\chi^2$ statistics when rejecting those stars in this region. For proper
motion data, $\chi^2/N_{\mathrm{eq}}$ decreases from 2.1 to 1.2. On the
contrary, for radial velocity data there is a little increase from 2.6 to
2.9. For the combined resolution a decrease from 2.4 to 1.3 was found. It
seems indicate that there is a better fit of the velocity field to the
model when rejecting this region. However, the value of
$\chi^2/N_{\mathrm{eq}}$ obtained is still larger than the derived in the
simulations (see \ref{appendix2}). This could be due by an underestimation
in the error in the photometric distances and/or radial velocities for our
stars.

Recently, Palou\u{s} (\cite{Palous}) used Hipparcos data to estimate the
second-order terms in the expansion of the velocity field of young stars
around the Sun. He concluded that these terms are always of low
significance and do not significantly alter the values obtained for the
first order derivatives (Oort constants). It cannot be ruled out that
higher order terms might account for the irregularities observed,
particularly the discrepancies appearing when some regions in $l$ are
omitted. However, the small number of stars available at large distances
does not allow us to avoid the large correlations between variables when
second order terms are included. We believe that the peculiar motions in
certain specific regions are largely responsible for the discrepancies
observed in the derivation of the $A$ Oort constant.

Likewise, we confirmed that the discrepancy on $V_\odot$ does not
disappear when eliminating the stars of the Orion region, neither when
eliminating stars from other regions. Again, from our simulations we
discarded the irregular distribution of the stars in the $X$-$Y$ plane as
an explanation for this discrepancy, so again the most probably
explanation is the departure from the adopted model. Finaly we would to
comment that the difference present in the derived $K$ value is a simple
consequence of the fact that this parameter is poorly determined when only
proper motion equations of stars with small galactic latitude are
considered.

Following we present the correlation matrices obtained for the radial
velocities, proper motions and combined solutions given in Table
\ref{tab.constants.vr.mp.vr+mp.Oort}:\\

\noindent
{\it i). Radial velocity solution:}
\begin{center}
{\small
\begin{tabular}{rrrrrrl}
  $A$   &   $C$   &   $K$   & $U_\odot$ & $V_\odot$ & $W_\odot$ &  \\
   1.00 & $-$0.12 & $-$0.07 &  $-$0.18  &  $-$0.15  &  $-$0.12  & $A$ \\ 
        &    1.00 & $-$0.03 &     0.05  &  $-$0.04  &  $-$0.00  & $C$ \\
        &         &    1.00 &  $-$0.11  &  $-$0.06  &  $-$0.09  & $K$ \\
        &         &         &     1.00  &  $-$0.08  &  $-$0.04  & $U_\odot$ \\
        &         &         &           &     1.00  &  $-$0.05  & $V_\odot$ \\
        &         &         &           &           &     1.00  & $W_\odot$ \\ 
\end{tabular}
}
\end{center}

\noindent
{\it ii). Proper motion solution:}
\begin{center}
{\small
\begin{tabular}{rrrrrrrl}
  $A$   &   $B$   &   $C$   &   $K$     & $U_\odot$ & $V_\odot$ & $W_\odot$ &  \\
   1.00 & $-$0.14 &    0.11 &  $-$0.06  &     0.05  &     0.07  &     0.06  & $A$ \\ 
        &    1.00 &    0.14 &     0.03  &  $-$0.02  &  $-$0.15  &  $-$0.01  & $B$ \\
        &         &    1.00 &     0.04  &  $-$0.22  &     0.10  &     0.01  & $C$ \\
        &         &         &     1.00  &  $-$0.04  &  $-$0.08  &     0.24  & $K$ \\
        &         &         &           &     1.00  &     0.10  &     0.04  & $U_\odot$ \\
        &         &         &           &           &     1.00  &     0.05  & $V_\odot$ \\ 
        &         &         &           &           &           &     1.00  & $W_\odot$ \\ 
\end{tabular}
}
\end{center}

\noindent
{\it iii). Combined solution:}
\begin{center}
{\small
\begin{tabular}{rrrrrrrl}
  $A$   &   $B$   &   $C$   &   $K$     & $U_\odot$ & $V_\odot$ & $W_\odot$ &  \\
   1.00 & $-$0.09 &    0.02 &  $-$0.07  &     0.07  &     0.08  &     0.01  & $A$ \\ 
        &    1.00 &    0.11 &     0.01  &  $-$0.02  &  $-$0.09  &  $-$0.01  & $B$ \\
        &         &    1.00 &     0.02  &  $-$0.10  &     0.02  &  $-$0.01  & $C$ \\
        &         &         &     1.00  &  $-$0.08  &  $-$0.07  &     0.01  & $K$ \\
        &         &         &           &     1.00  &     0.01  &     0.01  & $U_\odot$ \\
        &         &         &           &           &     1.00  &     0.05  & $V_\odot$ \\ 
        &         &         &           &           &           &     1.00  & $W_\odot$ \\ 
\end{tabular}
}
\end{center}

As can be seen, the correlations are small in all cases, constating that
they are not responsible of the discrepancies between proper motions and
radial velocities. We also realized that the combined solution presents
the smallest correlations.

The information that can be derived from the $\chi^2$ statistics provides
us further arguments to favour the combined solution in the analysis
performed in the following sections. As can be seen in Table
\ref{tab.constants.vr.mp.vr+mp.Oort}, the worst values for the fraction
$\chi^2/N_{\mathrm{eq}}$ are obtained in the radial velocity solution,
effect that is not reflected in our simulations. Two aspects can
contribute to this fact: an underestimation of the observational errors in
radial velocities -- which can not be rule out since this parameter is
very difficult to obtain for very hot stars -- or an error in the adoption
of the shape and size of the velocity dispersion ellipsoid. Both aspects
will be discussed hereafter but the only way to fully analyze the cosmic
dispersion velocity ellipsoid is through the residual analysis of the
combined solution. Henceforth, we will work hereafter with the combined
solution solving Eqs. (\ref{eq.vr}), (\ref{eq.vl}) and (\ref{eq.vb})
simultaneously, so taking into account all the information available.

\subsubsection{Cosmic dispersion}\label{cosmic}

Our residuals in Eqs. (\ref{eq.vr}), (\ref{eq.vl}) and (\ref{eq.vb})
include the observational errors, the residual velocity of the stars
(cosmic dispersion) and possible departures from the adopted linear model.
An optimal solution to characterize and separate the contribution of the
cosmic dispersion may be to adopt it from independent works such as Wielen
(\cite{Wielen}), Lacey (\cite{Lacey}), Asiain et al. (\cite{Asiain et
al.3}), among others, who analitically or empirically evaluated the
increase of the stellar velocity dispersion with age (disk heating).
Discrepancies are present in the first 100-150 Myr (our working domain),
where the population is not relaxed.

Working in the opposite sense, we can assume that the linear model adopted
is approximately correct and that the observational errors are well
estimated. With this hypothesis, and considering the interval 100 $< R
\le$ 600 pc to minimize both the possible departures from the lineal model
and the observational errors, we solved the combined solution for
different age intervals using an iterative process up to convergence. In
all cases one interation is sufficient and the results do not depend on
the adopted initial values. The results are presented in Table
\ref{tab.cosmica}.

\begin{table}[top]
   \caption{Standard deviation of the observational errors and cosmic
            dispersion for several subsamples divided in age intervals in
            the distance interval 100 $< R \le$  600 pc, both expressed in
            the galactic heliocentric coordinates system.
            $\sigma_{\mathrm{cos}} = 
            (\sigma_U^2+\sigma_V^2+\sigma_W^2)^{1/2}$: (1): this work,
            (2): Wielen (1977) analytical approximation $\sigma (t)^n =
            \sigma_o^n + C_v \,  t$ with $n = 2$ and $C_v = 6 \cdot 
            10^{-7}$ (km s$^{-1}$)$^2$ yr$^{-1}$ and $\sigma_o =$ 10 km
            s$^{-1}$. Units: km s$^{-1}$.} 
   \label{tab.cosmica}
\begin{tabular}{ccccc}
\hline
Age &($\sigma_{\epsilon_U}, \sigma_{\epsilon_V}, \sigma_{\epsilon_W})$ 
                                 & $(\sigma_U, \sigma_V, \sigma_W$)$_{\mathrm{cos}}$ 
                                 & $\sigma_{\mathrm{cos}}$
                                 & $\sigma_{\mathrm{cos}}$ \\
(Myr)&&&(1)&(2)\\
\hline
0 - 30   & (3.5, 3.0, 2.3) & (7.9, 7.2, 4.3)  & 11.5 & 10.4 \\
30 - 60  & (3.1, 3.4, 1.8) & (6.2, 7.5, 4.4)  & 10.7 & 11.3 \\
60 - 90  & (3.2, 3.2, 2.0) & (7.5, 8.8, 4.5)  & 12.4 & 12.0 \\
90 - 120 & (3.2, 3.1, 2.1) & (10.9, 9.6, 6.4) & 15.9 & 12.8 \\
$>$ 120  & (3.2, 3.2, 2.9) & (10.8,10.0, 5.5) & 15.7 & 13.8 \\
\hline
\end{tabular}
\end{table}

We must point out that the values obtained for the Oort parameters and the
solar motion components are practically independent of the choice of the
cosmic dispersion values, the differences in these parameters being always
smaller than 0.5 km s$^{-1}$ kpc$^{-1}$ or 0.5 km s$^{-1}$ respectively.
Conversely, the adopted values for the cosmic dispersion will directly
modify the $\chi^2$ statistic. With this in mind, and the fact that the
values derived for the cosmic dispersion are coherent with Wielen's work
(see Table \ref{tab.cosmica}), we proceded to solve combined solutions at
different age intervals using the above explained iterative process for
the interval 100 $ < R \le $ 600 pc. For the interval 600 $ < R \le $ 2000
pc and 100 $ < R \le $ 2000 pc we use as cosmic dispersion the values
obtained for 100 $< R \le $ 600 pc, that is, assuming isothermality. This
process give us the oportunity to use the $\chi^2$ statistic to evaluate
the over/underestimation of the observational errors or the departure from
the adopted linear model. The results are summarized in Table
\ref{tab.constants.Oort} and discussed in the following sections.

\begin{table*}[top]
   \caption{Oort constants and residual solar motion as a function of
            distance and age. Units: Age in Myr; $A$, $B$, $C$, $K$ in km
            s$^{-1}$ kpc$^{-1}$; $U_\odot$, $V_\odot$, $W_\odot$, $\sigma$
            in km s$^{-1}$. $\chi^2/N_{\mathrm{eq}}$ is the value of
            $\chi^2$ divided by the number of equations. $N$ is the number
            of stars (sample 1 + sample 2).}
   \label{tab.constants.Oort}
\begin{tabular}{crrrrrrrccc}
\hline
Age           &        $A$      &        $B$        &        $C$       &        $K$
              &     $U_\odot$   &     $V_\odot$     &     $W_\odot$    & $\sigma$ & $\chi^2/N_{\mathrm{eq}}$ &  $N$      \\
\hline
\hline
\multicolumn{11}{c}{100 $< R \le$ 600 pc} \\
\hline
\hline
0 - 30        &   5.7$_{(1.4)}$ & $-$20.7$_{(1.4)}$ &    5.2$_{(1.4)}$ &    7.1$_{(1.4)}$
              &   8.1$_{(0.5)}$ &    14.5$_{(0.4)}$ &    6.4$_{(0.3)}$ &  6.02    &  1.06    &  361 +  289 \\
30 - 60       &   7.6$_{(1.5)}$ & $-$14.5$_{(1.4)}$ &    9.5$_{(1.6)}$ &    4.0$_{(1.7)}$
              &  11.6$_{(0.4)}$ &    14.6$_{(0.5)}$ &    7.4$_{(0.3)}$ &  5.94    &  0.95    &  359 +  266 \\
$< 60$        &   6.3$_{(1.1)}$ & $-$18.5$_{(1.0)}$ &    5.9$_{(1.1)}$ &    5.1$_{(1.1)}$
              &   9.8$_{(0.3)}$ &    14.4$_{(0.3)}$ &    6.9$_{(0.2)}$ &  6.12    &  1.01    &  720 +  555 \\
60 - 90       &  10.5$_{(2.1)}$ & $-$13.6$_{(2.0)}$ &    5.9$_{(2.1)}$ & $-$5.4$_{(2.3)}$
              &  12.4$_{(0.5)}$ &    13.8$_{(0.6)}$ &    6.8$_{(0.4)}$ &  6.72    &  1.14    &  245 +  183 \\
$> 60$        &  11.8$_{(1.5)}$ & $-$11.0$_{(1.4)}$ & $-$0.9$_{(1.5)}$ & $-$3.5$_{(1.7)}$
              &  12.0$_{(0.4)}$ &    13.2$_{(0.4)}$ &    6.7$_{(0.2)}$ &  8.07    &  1.09    &  932 +  654 \\
$> 90$        &  11.9$_{(2.0)}$ &  $-$9.4$_{(1.8)}$ & $-$4.6$_{(2.0)}$ & $-$1.9$_{(2.2)}$
              &  11.8$_{(0.4)}$ &    12.9$_{(0.4)}$ &    6.6$_{(0.3)}$ &  8.30    &  1.09    &  687 +  471 \\
\hline
All           &   8.8$_{(0.8)}$ & $-$14.2$_{(0.7)}$ &    1.5$_{(0.8)}$ &    0.5$_{(0.9)}$
              &  11.2$_{(0.2)}$ &    13.0$_{(0.2)}$ &    6.7$_{(0.1)}$ &  7.24    &  1.06    & 2970 + 1596 \\
\hline
\hline
\multicolumn{11}{c}{600 $< R \le$ 2000 pc} \\
\hline
\hline
0 - 30        &  13.3$_{(0.7)}$ & $-$11.7$_{(0.7)}$ & $-$0.3$_{(0.7)}$ & $-$2.6$_{(0.7)}$
              &   8.0$_{(0.8)}$ &    12.9$_{(0.8)}$ &    7.9$_{(0.5)}$ & 10.39    &  1.95    &  285 +  204 \\
30 - 60       &   9.1$_{(1.7)}$ & $-$10.8$_{(1.8)}$ & $-$3.4$_{(1.8)}$ & $-$0.8$_{(1.8)}$
              &  15.0$_{(1.8)}$ &    10.4$_{(1.7)}$ &    8.9$_{(1.1)}$ & 11.57    &  2.29    &   81 +   56 \\
$< 60$        &  12.7$_{(0.6)}$ & $-$11.7$_{(0.7)}$ & $-$0.6$_{(0.7)}$ & $-$2.5$_{(0.6)}$
              &   9.1$_{(0.8)}$ &    12.5$_{(0.7)}$ &    8.0$_{(0.5)}$ & 10.91    &  2.05    &  366 +  260 \\
\hline
All           &  13.0$_{(0.7)}$ & $-$12.1$_{(0.7)}$ &    0.5$_{(0.8)}$ & $-$2.9$_{(0.6)}$
              &   9.0$_{(0.8)}$ &    13.4$_{(0.7)}$ &    8.3$_{(0.5)}$ & 11.77    &  1.87    &  449 +  308 \\
\hline
\hline
\multicolumn{11}{c}{100 $< R \le$ 2000 pc} \\
\hline
\hline
0 - 30        &  12.9$_{(0.6)}$ & $-$13.0$_{(0.6)}$ & $ $0.5$_{(0.6)}$ & $-$1.7$_{(0.5)}$
              &   8.6$_{(0.4)}$ &    13.4$_{(0.4)}$ &    6.7$_{(0.3)}$ &  7.85    &  1.51    &  646 +  493 \\
30 - 60       &   9.6$_{(1.0)}$ & $-$13.2$_{(1.0)}$ & $ $2.0$_{(1.0)}$ & $-$0.2$_{(1.0)}$
              &  11.7$_{(0.4)}$ &    13.4$_{(0.5)}$ &    7.4$_{(0.3)}$ &  6.85    &  1.19    &  440 +  322 \\
$< 60$        &  12.0$_{(0.5)}$ & $-$13.0$_{(0.5)}$ & $ $0.7$_{(0.5)}$ & $-$1.5$_{(0.5)}$
              &  10.0$_{(0.3)}$ &    13.4$_{(0.3)}$ &    7.1$_{(0.2)}$ &  7.52    &  1.34    & 1086 +  815 \\
$> 60$        &  11.1$_{(1.4)}$ & $-$12.2$_{(1.3)}$ & $-$1.8$_{(1.3)}$ & $-$5.1$_{(1.5)}$
              &  11.8$_{(0.4)}$ &    13.2$_{(0.4)}$ &    6.8$_{(0.2)}$ &  8.45    &  1.19    &  981 +  676 \\
\hline
All           &  11.8$_{(0.4)}$ & $-$12.3$_{(0.4)}$ &    0.4$_{(0.4)}$ & $-$2.0$_{(0.4)}$
              &  11.0$_{(0.2)}$ &    12.9$_{(0.2)}$ &    6.8$_{(0.1)}$ &  7.77    &  1.17    & 3419 + 1904 \\
\hline
\end{tabular}
\end{table*}

\subsection{Large scale outline of the local galactic kinematics}

An initial overview of the kinematics of young stars in the solar vicinity
is given when considering all the stars in the interval 100 $< R \le$ 2000
pc. The solar motion relative to the stellar group considered was found to
be:
\begin{equation}
  (U_\odot, V_\odot, W_\odot) = (11.0, 12.9, 6.8) \pm (0.2, 0.2, 0.1)
  {\mathrm{\; km \; s}}^{-1}
\end{equation}
On the other hand, we found the following values of the Oort constants,
which are dominated by a pure differential galactic rotation:
\begin{eqnarray}
A & = & 11.8 \pm 0.4\; {\rm \; km \; s}^{-1} {\rm \; kpc}^{-1} \nonumber \\
B & = & -12.3 \pm 0.4\; {\rm \; km \; s}^{-1} {\rm \; kpc}^{-1} \nonumber \\
C & = & 0.4 \pm 0.4\; {\rm \; km \; s}^{-1} {\rm \; kpc}^{-1} \nonumber \\
K & = & -2.0 \pm 0.4\; {\rm \; km \; s}^{-1} {\rm \; kpc}^{-1}
\end{eqnarray}

The kinematic distortion produced by the Gould Belt in the solar
neighbourhood can be removed when considering only those stars with 600 $< R
\le$ 2000 pc. Then, we found a solar motion:
\begin{equation}
  (U_\odot, V_\odot, W_\odot) = (9.0, 13.4, 8.3) \pm (0.8, 0.7, 0.5)
  {\mathrm{\; km \; s}}^{-1}
\end{equation}
in perfect agreement with the classic value, the changes with respect to
the former solution being partly produced by the presence of moving groups
among young stars (Asiain et al. \cite{Asiain et al.2}). From
\ref{appendix2} we expect these values to be slightly underestimated due
to a bias of $\approx -$(0.3-0.4) km s$^{-1}$. The Oort constants were
found to be:
\begin{eqnarray}
A & = & 13.0 \pm 0.7\; {\rm \; km \; s}^{-1} {\rm \; kpc}^{-1} \nonumber \\ 
B & = & -12.1 \pm 0.7\; {\rm \; km \; s}^{-1} {\rm \; kpc}^{-1} \nonumber \\ 
C & = & 0.5 \pm 0.8\; {\rm \; km \; s}^{-1} {\rm \; kpc}^{-1} \nonumber \\ 
K & = & -2.9 \pm 0.6\; {\rm \; km \; s}^{-1} {\rm \; kpc}^{-1} 
\end{eqnarray}
From \ref{appendix2}, a significant bias is only expected for $B$, with an
underestimation (in absolute value) of about 0.8 km s$^{-1}$ kpc$^{-1}$.
Therefore, its value might be $B \approx -12.9 \pm 0.7$ km s$^{-1}$
kpc$^{-1}$. These $A$ and $B$ Oort constant values are in good agreement
with the results obtained by Lindblad et al. (\cite{Lindblad et al.2}).
Using Hipparcos data they found $A = 13.7 \pm 1.0$ km \mbox{s$^{-1}$}
\mbox{kpc$^{-1}$} and $B = -13.6 \pm 0.8 $ km \mbox{s$^{-1}$}
\mbox{kpc$^{-1}$} from a sample of O and B stars with $R \le$ 2000 pc
outside the Gould Belt. Feast \& Whitelock (\cite{Feast et al.1}) reported
$A = 14.8 \pm 0.8$ km \mbox{s$^{-1}$} \mbox{kpc$^{-1}$} and $B = -12.4 \pm
0.6 $ km \mbox{s$^{-1}$} \mbox{kpc$^{-1}$} from a sample of Cepheid stars
(at distances up to 5000 pc) with Hipparcos proper motions and distance
calibration. Using a similar sample (also with an Hipparcos distance
calibration), Feast et al. (\cite{Feast et al.2}) found $A = 15.1 \pm 0.3$
km \mbox{s$^{-1}$} \mbox{kpc$^{-1}$} from radial velocities. This tendency
to obtain lower values for the $A$ Oort constant when the distance horizon
of the sample is approached is confirmed in the results presented in Table
\ref{tab.constants.Oort}. Using stars with 100 $< R \le$ 600 pc not
belonging to the Gould Belt (age larger than 90 Myr), we found an $A$
constant of 11.9 $\pm$ 2.0 km s$^{-1}$ kpc$^{-1}$, roughly 1 km s$^{-1}$
kpc$^{-1}$ less than that obtained for all the stars with 600 $< R \le$
2000 pc. When comparing this result with the $A = 11.3 \pm 1.1$ km
s$^{-1}$ kpc$^{-1}$ derived by Hanson (\cite{Hanson}) using proper motions
for approximately 60\,000 nearby faint stars ($R <$ 1000 pc and a
photographic magnitude 16 $< m_{\mathrm{pg}} <$ 17) from the Lick Northern
Proper Motion (NPM) program, we found a good coherence, although his
sample was composed basically of F2-K0 stars. Recently, Olling \&
Merrifield (\cite{Olling et al.}), using a mass model which includes the
interstellar gas component, derived the variation in the Oort constants as
a function of the galactocentric distance. The authors explained the
discrepancies between the $A$ values derived by Hanson (\cite{Hanson}) and
Feast et al. (\cite{Feast et al.2}) as being produced by the different
mean galactocentric distances of the two samples. However, the authors
admitted a potential source of error in their analysis due to the
assumption of azimuthal symmetry in the orbital structure of the Galaxy.  

Our nearly null values of $C$ and $K$ constants when considering all the
stars in the distance intervals 100 $< R \le$ 2000 pc and 600 $< R \le$
2000 pc are in good agreement with a pure differential galactic rotation.
CTG found a value of $K \approx -$(1-2) km \mbox{s$^{-1}$}
\mbox{kpc$^{-1}$} for stars with $R <$ 1500 pc, very similar to ours.
Nevertheless, for the $C$ constant they found a clearly negative value for
B6-A0 stars: $C = -8.8 \pm 1.1 $ km \mbox{s$^{-1}$} \mbox{kpc$^{-1}$}.
This negative value of $C$ seems to be corroborated by Mestres
(\cite{Mestres}), who found $C = -4.8 \pm 1.2 $ km \mbox{s$^{-1}$}
\mbox{kpc$^{-1}$} for those stars with 400 $< R \le$ 1500 pc. More
recently, Lindblad et al. (\cite{Lindblad et al.2})  found $C = 0.8 \pm
1.1 $ km \mbox{s$^{-1}$} \mbox{kpc$^{-1}$} and $K = -1.1 \pm 0.8$ km
\mbox{s$^{-1}$} \mbox{kpc$^{-1}$} from their sample of O and B Hipparcos
stars outside the Gould Belt. This last result is in very good agreement
with ours.

To evaluate the goodness of our fit we can look at the $\chi^2$
statistics. We found values 1.9-2.3 for $\chi^2/N_{\mathrm{eq}}$,
depending on the age interval considered. For a moderately good fit we
would expect a value $\chi^2/N_{\mathrm{eq}} \approx 1$. We think the
difference to be produced by an underestimation in the errors in the
photometric distances and radial velocities for far stars.

This overall view is changed considerably when we divide our sample of
stars into age and distance groups. The most characteristic change when we
study the system of the nearest and youngest stars is the kinematic
signature of the Gould Belt, i.e. the appearence of a non-null value of
the $K$ Oort constant and the peculiar behaviour of the other Oort
constants.

\subsection{Local irregularities: the kinematic effects of the Gould
Belt}\label{sect.gould}

\begin{figure}
  \resizebox{\hsize}{!}{\includegraphics{9113.f11}}
  \caption{Variation of Oort parameters plotted against the age for stars
           within 100 $< R \le$ 600 pc. Units are in km s$^{-1}$ kpc$^{-1}$.}
  \label{fig.oort}
\end{figure}

To study the kinematic characteristics of the Gould Belt, we present in
Table \ref{tab.constants.Oort} the results for those stars in our sample
with 100 $< R \le$ 600 pc. In Fig. \ref{fig.oort}, we show the variation
of the Oort parameters with age. In general, we observe a marked increase
in $A$ and $B$ values with age, and a decrease in $C$ and $K$, according
to the results obtained by Torra et al. (\cite{Torra et al.}).

A non-pure differential galactic rotation was found for the youngest group
of stars, with $A = 5.7 \pm 1.4$ km s$^{-1}$ kpc$^{-1}$, $B = -20.7 \pm
1.4$ km s$^{-1}$ kpc$^{-1}$ and non-null values for $C$ and $K$.  The
tendency to obtain small $A$ and $B$ values for the youngest group is in
perfect agreement with the results obtained by Lindblad et al.
(\cite{Lindblad et al.2}) and Torra et al. (\cite{Torra et al.}), although
these autors even report negative $A$ values in the combined solution. In
agreement with these two studies, we confirmed that significant
differences do appear in the solutions using only radial velocities or
only proper motion equations ($A = 0.7 \pm 2.9$ km s$^{-1}$ kpc$^{-1}$ and
$A = 6.4 \pm 1.4$ km s$^{-1}$ kpc$^{-1}$, respectively), and as discussed
in the previous section, a possible explanation could be the departure of
some stellar groups from the adopted linear model. When we considered not
so young stars, $A$ and $B$ are closer to classic values ($A \approx 12$
km s$^{-1}$ kpc$^{-1}$ and $B \approx -9$ km s$^{-1}$ kpc$^{-1}$). From
\ref{appendix2} we can see that these differences are not expected to be
produced by any systematic bias. In this distance interval we only expect
an underestimation in $A$ of about 0.5 km s$^{-1}$ kpc$^{-1}$.

\begin{figure*}
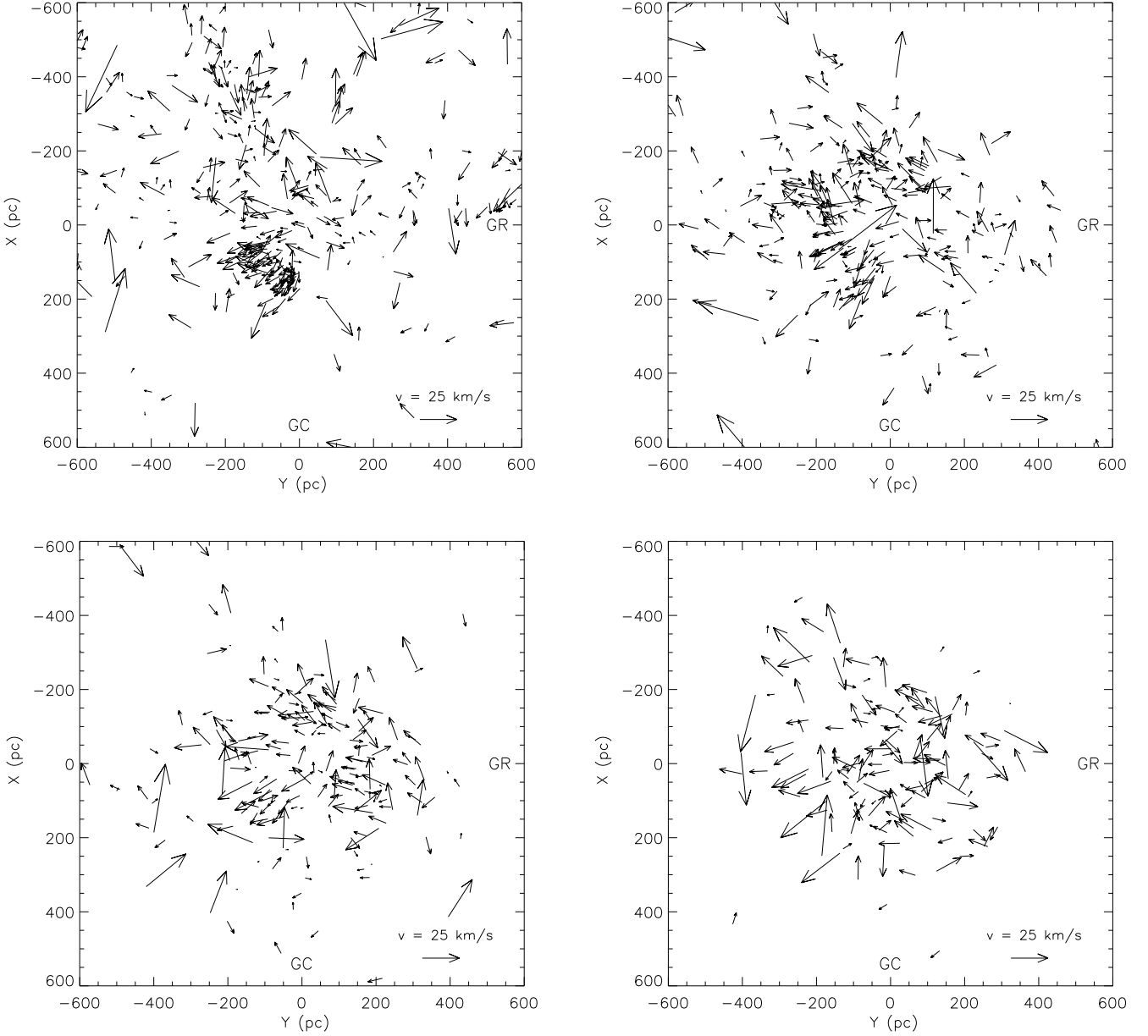

  \mbox{}
  \vspace{16.5cm}
  \includegraphics{9113a.f12}
  \includegraphics{9113b.f12}
  \includegraphics{9113c.f12}
  \includegraphics{9113d.f12}
  \caption[]{Residual heliocentric space velocity vectors projected on the
             galactic plane for O and B stars with an age $\tau \le$ 30
             Myr (top left), 30 $< \tau \le$ 60 Myr (top right), 60 $<
             \tau \le$ 90 Myr (bottom left) and 90 $< \tau \le$ 120 Myr
             (bottom right).}
  \label{fig.residu.oort}
\end{figure*}

For those stars with $\tau \le$ 60 Myr we found a clear positive value of
$K$-term: $K = 7.1 \pm 1.4$ km s$^{-1}$ kpc$^{-1}$ for $\tau \le$ 30 Myr,
and $K = 4.0 \pm 1.7$ km s$^{-1}$ kpc$^{-1}$ for 30 $< \tau \le$ 60 Myr.
On the other hand, for stars older than 60 Myr slightly negative values of
$K$ were found -- nearly compatible with a null value. Positive values of
$K$ were not found when we considered the distance interval 600 $< R \le$
2000 pc, independently of the age interval.

In the interval 100 $ < R \le $ 600 pc the obtained values for
$\chi^2/N_{\mathrm{eq}}$ were similar to those derived from the
simulations (see \ref{appendix2}), around 1.0. From the coherent values of
the cosmic dispersion obtained, we can conclude that the velocity field of
our stars fits the lineal model proposed and the errors are well
estimated.

The variations in Oort parameters as a function of age allow us to infer
an estimation of the age of the Gould Belt. As we have seen, when
considering stars with $100 < R \le 600$ pc a nearly pure differential
galactic rotation was only found for stars with an age greater than 90
Myr. In the age interval 60-90 Myr, rather low values of $A$ and $B$ and a
high value of $C$ were still derived. So, we conclude that the age of the
Gould Belt derived from the kinematic behaviour of the stars is in perfect
agreement with that derived in Sect. \ref{sect.age} from the analysis of
their spatial distribution.

\begin{figure*}
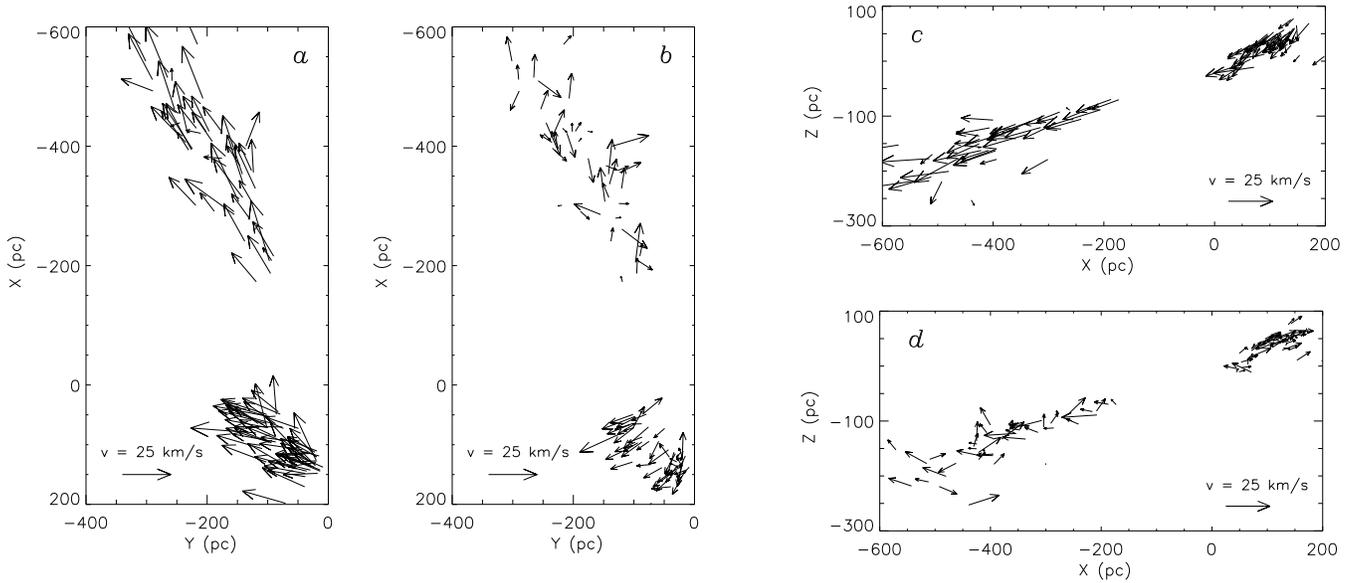

  \mbox{}
  \vspace{8cm}
  \includegraphics{9113a.f13}
  \includegraphics{9113b.f13}
  \includegraphics{9113c.f13}
  \includegraphics{9113d.f13}
  \caption[]{Heliocentric space velocity vectors ($a$,$c$) and residual
             velocity vectors ($b$,$d$) projected on the galactic plane
             ($X$-$Y$) and the meridional plane ($X$-$Z$) for stars
             younger than 30 Myr belonging to the complexes Sco-Cen and
             Ori OB1.}
  \label{fig.residu.oort.assoc}
\end{figure*}

\begin{table*}[top]
   \caption{Oort parameters and residual solar motion for stars with 100
            $< R \le$ 600 pc and $\tau \le$ 30 Myr when excluding those stars
            belonging to the complexes Sco-Cen and Ori OB1, following the
            member lists provided by Brown et al. (\cite{Brown et al.1};
            Ori OB1 association) and de Zeeuw et al. (\cite{de Zeeuw et
            al.}; Sco-Cen complex). Units: $A$, $B$, $C$, $K$ in km
            s$^{-1}$ kpc$^{-1}$; $U_\odot$, $V_\odot$, $W_\odot$, $\sigma$ 
            in km s$^{-1}$. $\chi^2/N_{\mathrm{eq}}$ is the value of
            $\chi^2$ divided by the number of equations. $N$ is the number
            of stars (sample 1 + sample 2).}
   \label{tab.constants.assoc.Oort}
\begin{tabular}{crrrrrrrccc}
\hline
Excluded            &        $A$      &        $B$        &        $C$       &        $K$
                    &     $U_\odot$   &     $V_\odot$     &     $W_\odot$    & $\sigma$ & $\chi^2/N_{\mathrm{eq}}$ &  $N$      \\
\hline
\hline
None                &   5.7$_{(1.4)}$ & $-$20.7$_{(1.4)}$ &    5.2$_{(1.4)}$ &    7.1$_{(1.4)}$
                    &   8.1$_{(0.5)}$ &    14.5$_{(0.4)}$ &    6.4$_{(0.3)}$ &  6.02    &  1.06    &  361 +  289 \\
Sco-Cen             &   6.9$_{(1.6)}$ & $-$19.7$_{(1.6)}$ &    4.7$_{(1.6)}$ &    5.8$_{(1.6)}$
                    &   8.5$_{(0.6)}$ &    13.9$_{(0.5)}$ &    6.2$_{(0.3)}$ &  6.39    &  1.18    &  305 +  238 \\
Ori OB1             &   6.1$_{(1.6)}$ & $-$20.7$_{(1.6)}$ &    5.3$_{(1.6)}$ &    7.3$_{(1.6)}$
                    &   8.0$_{(0.5)}$ &    14.6$_{(0.4)}$ &    6.6$_{(0.3)}$ &  6.14    &  1.10    &  315 +  251 \\
Both complexes      &   7.2$_{(1.8)}$ & $-$19.7$_{(1.8)}$ &    4.9$_{(1.9)}$ &    6.0$_{(1.9)}$
                    &   8.4$_{(0.6)}$ &    14.0$_{(0.6)}$ &    6.4$_{(0.3)}$ &  6.60    &  1.26    &  258 +  200 \\
\hline
\end{tabular}
\end{table*}

In Fig. \ref{fig.residu.oort} we show the residual space velocity vectors
for each star projected on the galactic plane, after subtracting the solar
motion and the galactic rotation found in the resolution for all the stars
in the distance interval 600 $< R \le$ 2000 pc, classified in different
age groups. From this figure, it is evident that to model the expansion of
the Gould Belt system as an expansion from a point (Olano \cite{Olano}) or
a line oriented in the direction $l = 45\degr \rightarrow 225\degr$ (CTG)
is not a good approximation. In the same figure we have identified certain
clumps associated with Lac OB1, Cep OB2, Cas-Tau, Per OB2, Col 121, Vel
OB2, Tr 10 and Sco-Cen. All these associations, except Cep OB2, possibly
belong to the Gould Belt (Comer\'on \cite{Comeron1}; P\"oppel
\cite{Poppel}). In the youngest group (top left panel: stars with $\tau
\le$ 30 Myr) we observe the clear residual motion of the Sco-Cen complex.
This complex is composed of three associations -- Upper Scorpius, US;
Upper Centaurus Lupus, UCL; and Lower Centaurus Crux, LCC -- at a distance
of about 120-145 pc, in the region of positive $X$ and negative $Y$. The
Ori OB1 association shows a smaller mean residual motion, as will be
discussed later. It is composed of several subgroups, situated in a range
of distances of about 340-510 pc, in the direction of $l \approx
200$-$210\degr$.

Sco-Cen and Ori OB1 are the two main complexes in the Gould Belt, and it
is particularly interesting to study their motion and influence on the
velocity field separately. This study might clarify whether the Gould Belt
is a casual arrangement of OB associations or a structure with a common
origin. In Fig. \ref{fig.residu.oort.assoc} we show the heliocentric
velocity field and the residual velocity field -- computed as in Fig.
\ref{fig.residu.oort} -- of the stars belonging to the Sco-Cen and Ori OB1
complexes on the $X$-$Y$ and $X$-$Z$ galactic planes. To select these
stars, we used the member lists provided by Brown et al. (\cite{Brown et
al.1}; Ori OB1 association) and de Zeeuw et al. (\cite{de Zeeuw et al.};
Sco-Cen complex). It should be noted that we have detected -- especially
in the case of Sco-Cen -- some additional stars whose location and
kinematics are compatible with membership to these associations. We
confirmed a high residual velocity field for Sco-Cen, which is moving away
from the Sun: $(U, V, W)_{\mathrm{res}} = (4.1, -6.7, 2.2)$ km s$^{-1}$,
$(v_{\mathrm{r}}, v_{\mathrm{l}},
v_{\mathrm{b}})_{\mathrm{res}}\footnote{$v_{\mathrm{l}} = 4.741 R
\mu_{\mathrm{l}} \cos b$ and $v_{\mathrm{b}} = 4.741 R \mu_{\mathrm{b}}$}
= (7.1, -3.2, 0.6)$ km s$^{-1}$. In the case of the Ori OB1 association,
the mean residual motion found was smaller and practically null in the
radial direction: $(U, V, W)_{\mathrm{res}} = (-2.7, 3.1, 3.6)$ km
s$^{-1}$, $(v_{\mathrm{r}}, v_{\mathrm{l}}, v_{\mathrm{b}})_{\mathrm{res}}
= (0.2, -3.9, 3.8)$ km s$^{-1}$. These effects were confirmed when
deriving the Oort parameters for those stars with 100 $< R \le$ 600 pc and
younger than 30 Myr not identified as members of these complexes (see
Table \ref{tab.constants.assoc.Oort}). When Ori OB1 was excluded from the
calculation of Oort parameters, the value of the $K$-term was found to
rise only by 0.2 km s$^{-1}$ kpc$^{-1}$. In contrast, when Sco-Cen was
excluded a decrease of 1.3 km s$^{-1}$ kpc$^{-1}$ in the $K$ value was
found. In both cases the $A$, $B$ and $C$ Oort parameters change less than
about 1 km s$^{-1}$ kpc$^{-1}$. When both complexes were eliminated the
parameters obtained were very similar to those obtained earlier (all the
changes are within the error bars).

We conclude that these associations are not the only responsible for the
peculiar kinematics observed for the youngest stars in the solar
neighbourhood, attributed to the Gould Belt. So, other nearby associations
and field stars belonging to the Gould Belt have a great significance in
the determination of the Oort parameters.

\begin{figure}
  \resizebox{\hsize}{!}{\includegraphics{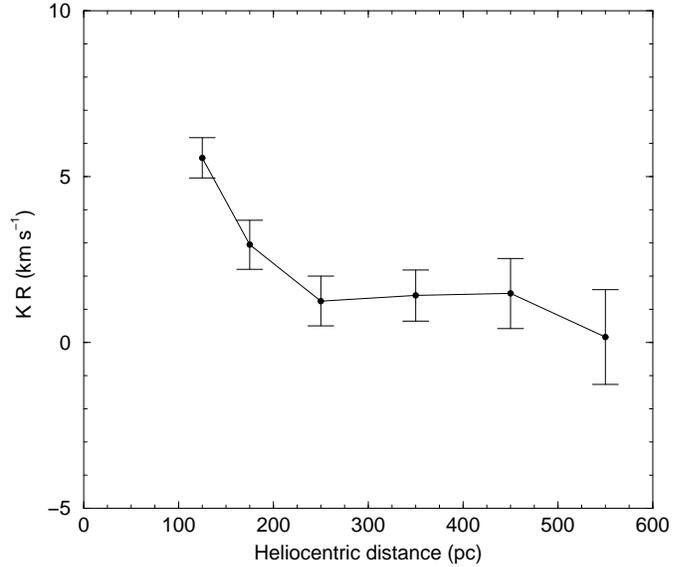}}
  \caption{Variation of $K R$ against the heliocentric distance for stars
           with an age smaller than 60 Myr.}
  \label{fig.k}
\end{figure}

\begin{figure*}
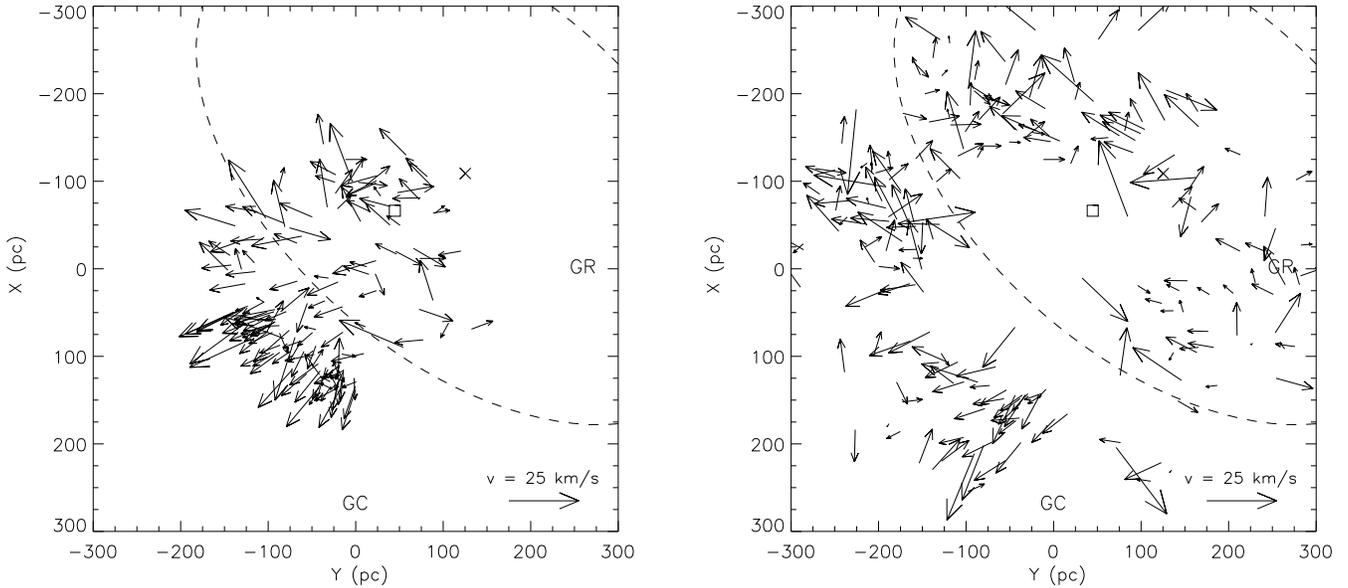

  \mbox{}
  \vspace{8cm}
  \includegraphics{9113a.f15}
  \includegraphics{9113b.f15}
  \caption[]{Residual heliocentric space velocity vectors projected on the
             galactic plane for O and B stars with an age $\tau \le$ 60
             Myr and 0 $< R \le$ 150 pc (left) and 150 $< R \le$ 300 pc
             (right). It is also drawn the Olano's (\cite{Olano}) Lindblad
             Ring (with a cross in its center) and the center of the Gould
             Belt (square) obtained by Comer\'on \& Torra (\cite{Comeron
             et al.1}). GC and GR indicate the galactic center and the
             galactic rotation directions respectively.}
  \label{fig.residu2.oort}
\end{figure*}

An attempt was made at analysing the expansion of the system as a function
of distance. As a first step, in Fig. \ref{fig.k} we present the variation
of the $K R$ product as a function of heliocentric distance for stars with
$\tau \le$ 60 Myr. The radial expansion disminishes rapidly with
increasing distance (for $R <$ 250 pc) and it does not extend further than
400 pc. At distances larger than 300 pc, only Per OB2 has a mean residual
motion away from the Sun. As discussed above, Ori OB1 has an almost null
radial residual motion. Even considering the solar motion proposed by
Dehnen \& Binney (\cite{Dehnen et al.}) -- $(U_\odot,V_\odot,W_\odot) =
(10.00, 5.25, 7.17)$ km s$^{-1}$ -- and the galactic rotation curve
obtained by Feast \& Whitelock (\cite{Feast et al.1}), we obtain a small
residual motion for this agreggate of $(U, V, W)_{\mathrm{res}} = (-1.2,
-2.8, 2.1)$ km s$^{-1}$, $(v_{\mathrm{r}}, v_{\mathrm{l}},
v_{\mathrm{b}})_{\mathrm{res}} = (1.7, 1.9, 2.8)$ km s$^{-1}$. To analyse
in more detail the expansion in the $R <$ 300 pc region, we show in Fig.
\ref{fig.residu2.oort} the residual velocity vectors projected on the
galactic plane for stars with $\tau \le$ 60 Myr (for clarity we divided
the sample in two different coronae: $R \le$ 150 pc and 150 $< R \le$ 300
pc) together with Olano's (\cite{Olano}) Lindblad Ring -- with a center
placed at $R = 166$ pc from the Sun in the direction $l = 131\degr$, and
semiaxes of 364 pc and 211 pc -- and the position of the center of the
Gould Belt proposed by Comer\'on \& Torra (\cite{Comeron et al.1}) -- $R =
80$ pc, $l = 146\degr$. In the first quadrant there is a lack of stars,
partially produced by the near high extinction structures related to the
Ophiuchus-Aquila complex (Vergely et al. \cite{Vergely et al.}), which
does not allow us to study the residual velocity field in this region. The
Cas-Tau complex is moving in the direction $l \approx$ 240\degr. So, as
shown in Figs. \ref{fig.residu.oort} and \ref{fig.residu2.oort} and by
Lindblad et al. (\cite{Lindblad et al.2}), the third and fourth galactic
quadrants contain the most significant structures accounting for the
expansion of the Gould Belt.

\begin{figure}
  \resizebox{\hsize}{!}{\includegraphics{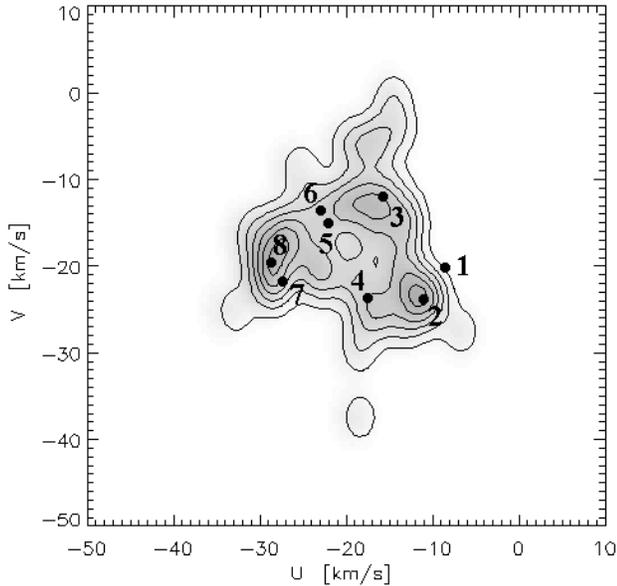}}
  \caption{Distribution of stars in the $U$-$V$ plane -- where a kernel
           estimator was used to indicate the lines of isocontours -- for
           stars with 225 $< l \le$ 285\degr, 100 $< R \le$ 300 pc and 30
           $< \tau \le$ 60 Myr. The filled circles correspond to the
           position of the clusters in Table \ref{tab.clusters}.}
  \label{fig.kernel}
\end{figure}

Whereas the fourth quadrant contains the extensively studied Sco-Cen
agreggate, a well-defined concentration of O and B stars with two
different residual motions is present in the region $225 \la l \la
285\degr$, mainly with distances in the interval $100 \la R \la 300$ pc
and ages between 30 and 60 Myr. To analyse these streams in detail we
present in Fig. \ref{fig.kernel} the distribution of these stars in the
$U$-$V$ plane, where a kernel estimator (Silverman \cite{Silverman}) was
used to indicate the isocontours. Over this figure, the mean $(U,V)$
heliocentric velocity components of the open clusters present in this
region derived from Hipparcos data by Robichon et al. (\cite{Robichon et
al.}) and the new kinematic structures recently identified by Platais et
al. (\cite{Platais et al.}) were also plotted. The stream placed at $(U,V)
= (-12,-23)$ km s$^{-1}$ shares the motion of the a Car (= HIP 45080)
cluster and IC 2602 open cluster, and it may also be related to the
Pleiades moving group substructures found by Asiain et al. (\cite{Asiain
et al.2}). The concentration observed at $(U,V) = (-28,-20)$ km s$^{-1}$
is associated with NGC 2451 A and Tr 10. The nature of NGC 2451 has been
under discussion for some time. According to R\"oser \& Bastian
(\cite{Roser et al.}), NGC 2451 can be divided into two different
entities. These authors named the closest of these the Puppis Moving Group
(PMG), the center of its distribution being clearly offset from the
nucleus of NGC 2451 by 1\degr. The distance to PMG was found to be 220 pc.
Carrier et al. (\cite{Carrier et al.}) also found two entities, at 198 and
358 pc respectively, from Geneva photometry and Hipparcos data. Two
well-defined peaks in parallax seem to reinforce the nature of open
cluster for these two entities, although the most distant is difficult to
distinguish from the field stars because both the parallax and proper
motion of its stars are close to those of field stars. On the other hand,
Tr 10 was identified as an intermediate age OB association by de Zeeuw et
al. (\cite{de Zeeuw et al.}), who found 23 members spread over
$\sim8\degr$ in the sky. Robichon et al. (\cite{Robichon et al.}) found 9
Hipparcos members of the cluster. The distance derived in both papers is
the same (365 pc). The $U$ velocity component found for the association is
the same as for the cluster ($U = -27.3$ km s$^{-1}$), but there is a
difference of 4 km s$^{-1}$ in the $V$ component ($V_\mathrm{assoc} =
-17.8$ km s$^{-1}$, $V_\mathrm{clu} = -21.9$ km s$^{-1}$).

We found that only 7 stars in this region (225 $< l \le 285\degr$, 100 $<
R \le$ 300 pc and 30 $< \tau \le$ 60 Myr) have been identified by the
above mentioned authors as members of open clusters or associations
(including the list of de Zeeuw et al. \cite{de Zeeuw et al.}), so a large
number of the remaining stars share the motion of these clusters and
associations and are spread over a large region, as observed in Fig.
\ref{fig.residu2.oort}. Finally, although IC 2391 and HR 3661 (= HIP
45189) seem to be isolated in Fig. \ref{fig.kernel}, stars sharing the
motion of these clusters were detected in our sample when the age interval
was changed to 60 $\le \tau \le$ 90 Myr. This is in agreement with the
estimated age for HR 3661 -- 100 Myr (Platais et al. \cite{Platais et
al.}) -- but not with the age of IC 2391 -- 30 Myr (Stauffer et al.
\cite{Stauffer et al.}).

\begin{table}[top]
   \caption{Galactic coordinates and heliocentric velocity components of
            the clusters shown in Fig. \ref{fig.kernel}. Units: $l$, $b$
            in degrees; $R$ in pc; $U$, $V$ and $W$ in km s$^{-1}$.}
   \label{tab.clusters}
\begin{tabular}{lcrcrrr}
\hline
Cluster           &  $l$  &   $b$   & $R$ &  $U$  &  $V$  &  $W$  \\
\hline
\hline
1. IC 2602$^2$  & 289.6 &  $-$4.9 & 152 &  $-$8 & $-$20 &  $-$0 \\
2. a Car$^1$    & 277.7 &  $-$7.6 & 132 & $-$11 & $-$24 &  $-$4 \\
3. NGC 2232$^2$ & 214.3 &  $-$7.7 & 325 & $-$16 & $-$12 & $-$11 \\
4. NGC 2516$^2$ & 273.9 & $-$15.9 & 346 & $-$17 & $-$24 &  $-$4 \\
5. HR 3661$^1$  & 266.9 &     3.4 & 174 & $-$22 & $-$15 &  $-$6 \\
6. IC 2391$^2$  & 270.4 &  $-$6.9 & 146 & $-$23 & $-$14 &  $-$7 \\
7. Tr 10$^2$    & 262.8 &     0.6 & 365 & $-$27 & $-$22 & $-$10 \\
8. NGC 2451$^2$ & 252.4 &  $-$6.8 & 189 & $-$29 & $-$20 & $-$14 \\
\hline
\hline
\multicolumn{7}{l}{$^1$ Platais et al. (\cite{Platais et al.})} \\
\multicolumn{7}{l}{$^2$ Robichon et al. (\cite{Robichon et al.})} \\
\end{tabular}
\end{table}

Further work will be necessary to confirm the existence of these streams
and to deal with their origin in the context of the various models
proposed for the Gould Belt. As an example, the fact that the $(U,V)$
motion of the stream around IC 2451 A is very similar to the motion of an
older (100-400 Myr) moving group independently detected by Figueras et al.
(\cite{Figueras et al.1}), Asiain et al. (\cite{Asiain et al.2}), Chereul
et al. (\cite{Chereul et al.}) and Sabas (\cite{Sabas}) raises interesting
questions. As a starting point, the vertical motions of these structures
should enable us to confirm or reject any relationship. In this way,
Comer\'on (\cite{Comeron2}) has recently reported a systematic gradient in
the vertical component of the velocity of those stars belonging to the
Gould Belt along the galactic plane, which while being subtle is
detectable in the Hipparcos astrometric data.

%
\section{Conclusions}

A sample of O- and B-type stars with Hipparcos astrometric data, radial
velocities and Str\"omgren photometry -- from which photometric distances
and ages were computed -- has been used to study the spatial distribution
and the kinematics of the young star system in the solar neighbourhood.
Several numerical simulations have allowed us to assess the robustness of
our methods and to evaluate the biases induced by the observational
constraints.

The spatial distribution of the youngest and nearest group of stars is
dominated by the presence of the Gould Belt. We found that this system
extended up to 600 pc from the Sun and has an orientation with respect to
the galactic plane of $i_{\mathrm{G}} = 16$-$22\degr$ and
$\Omega_{\mathrm{G}} = 275$-$295\degr$, depending on the distance and age
intervals considered. For $R \le$ 600 pc, roughly 60\% of stars younger
than 60 Myr belong to the Gould Belt.

In the region with $R >$ 600 pc, the stellar kinematics is dominated by
the differential galactic rotation, since the Oort constants were found to
be $A = 13.0 \pm 0.7$ km s$^{-1}$ kpc$^{-1}$, $B = -12.1 \pm 0.7$ km
s$^{-1}$ kpc$^{-1}$, $C = 0.5 \pm 0.8$ km s$^{-1}$ kpc$^{-1}$ and $K =
-2.9 \pm 0.6$ km s$^{-1}$ kpc$^{-1}$. In contrast, in the region with $R
\le$ 600 pc, the Gould Belt dominates the kinematics of the youngest stars
($\tau \leq$ 60 Myr), producing a decrease in the $A$ and $B$ Oort
constants ($A \approx$ 6-8 km s$^{-1}$ kpc$^{-1}$, $B \approx -$(21-14) km
s$^{-1}$ kpc$^{-1}$) and an increase in $C$ and $K$ ($C \approx$ 5-9 km
s$^{-1}$ kpc$^{-1}$, $K \approx$ 4-7 km s$^{-1}$ kpc$^{-1}$). This
peculiar kinematics was also found when those stars belonging to the
Sco-Cen and Ori OB1 complexes were eliminated. Therefore, these
associations are not the only responsible for these peculiarities, a
finding which seems to reinforce the suggestion by Guillout et al.
(\cite{Guillout et al.2}) that the Gould Belt is a disk-like rather than a
ring-like structure.

A perfect agreement has been obtained when estimating the age of the Gould
Belt system from the spatial distribution of the stars and from the study
of the variations in the Oort constants with age. Taking into account the
biases in the computation of individual photometric ages -- stellar
rotation and important uncertainties -- we estimated an age of the Gould
Belt inside the interval 30-60 Myr.

The study of the residual velocity field allowed us to estimate the cosmic
dispersion for stars younger that $\approx 150$ Myr. On the other hand,
this residual velocity field for the youngest stars cannot be explained as
an expansion from a point or a line. Moreover, the expansion motion
classically attributed to the Gould Belt seems to be due to the nearest
stars ($R \la$ 300 pc). In the region 300 $< R \le$ 600 pc, we found that
only Per OB2 has a clear residual motion away from the Sun. In the region
with 100 $\la R \la$ 300 pc and 225 $\la l \la 285\degr$, two streams of
stars with an age between 30 and 60 Myr have been found. One of these
streams shares the motion of a Car (= HIP 45080) and IC 2602, whereas the
other follows the motion of NGC 2451 A and Tr 10.

%
\begin{acknowledgements}

We thanked the anonymous referee for his comments and suggestions that
have improved the quality of this paper. This study has been suported by
the CICYT under contract ESP 97-1803 and by the PICS programme (CIRIT). DF
acknowledges the FRD grant of the Universitat de Barcelona (Spain).

\end{acknowledgements}

%

%
\appendix

\section{Simulations to check the structure analysis}\label{appendix1}

Simulations were performed to assess how incompleteness effects could
change our conclusions on the Gould Belt's structure parameters. The
critical questions to answer are:

\begin{itemize}

\item Due to observational constraints older stars ($\tau >$ 60 Myr) have
a small limiting distance ($R \approx$ 400 pc). If their spatial
distribution shows an inclined structure, as obtained for young and
distant stars, could our method, which does not take into account
incompleteness effects, be capable to detect it?

\item Related to the first point, for which scale height of the belts our
method looses its statistical capability?

\item Are the available number of stars enough to undertake this study?
Can simulations provide a realistic estimate of the errors in the derived
structure parameters?

\end{itemize}

To answer these questions simulated samples were built by considering the
following steps:

\begin{itemize}

\item From each real star we generated a pseudo-star with the same age,
visual magnitude and projected distance ($R \cos b$). Its galactic
longitude was randomly assigned and the distance to the galactic plane
($z$) simulated following an exponential distribution with scale height
$Z_0$.

\item Inside each age interval, the position of a fraction $q = 0.50$ of
the generated pseudo-stars were rotated an angle $i_{\mathrm{G}} = 20
\degr$ around the $Y$ axis (galactic rotation direction), that is adopting
$\Omega_{\mathrm{G}} = 270\degr$ for the ascending node of the Gould Belt.

\item The process was repeated to generate samples with $Z_0$ values
ranging from 40 to 80 pc (Mihalas and Binney \cite{Mihalas et al.} quoted
$Z_0 = 60$ pc for B-type stars).

\end{itemize}

\begin{figure*}
  \resizebox{\hsize}{!}{\includegraphics{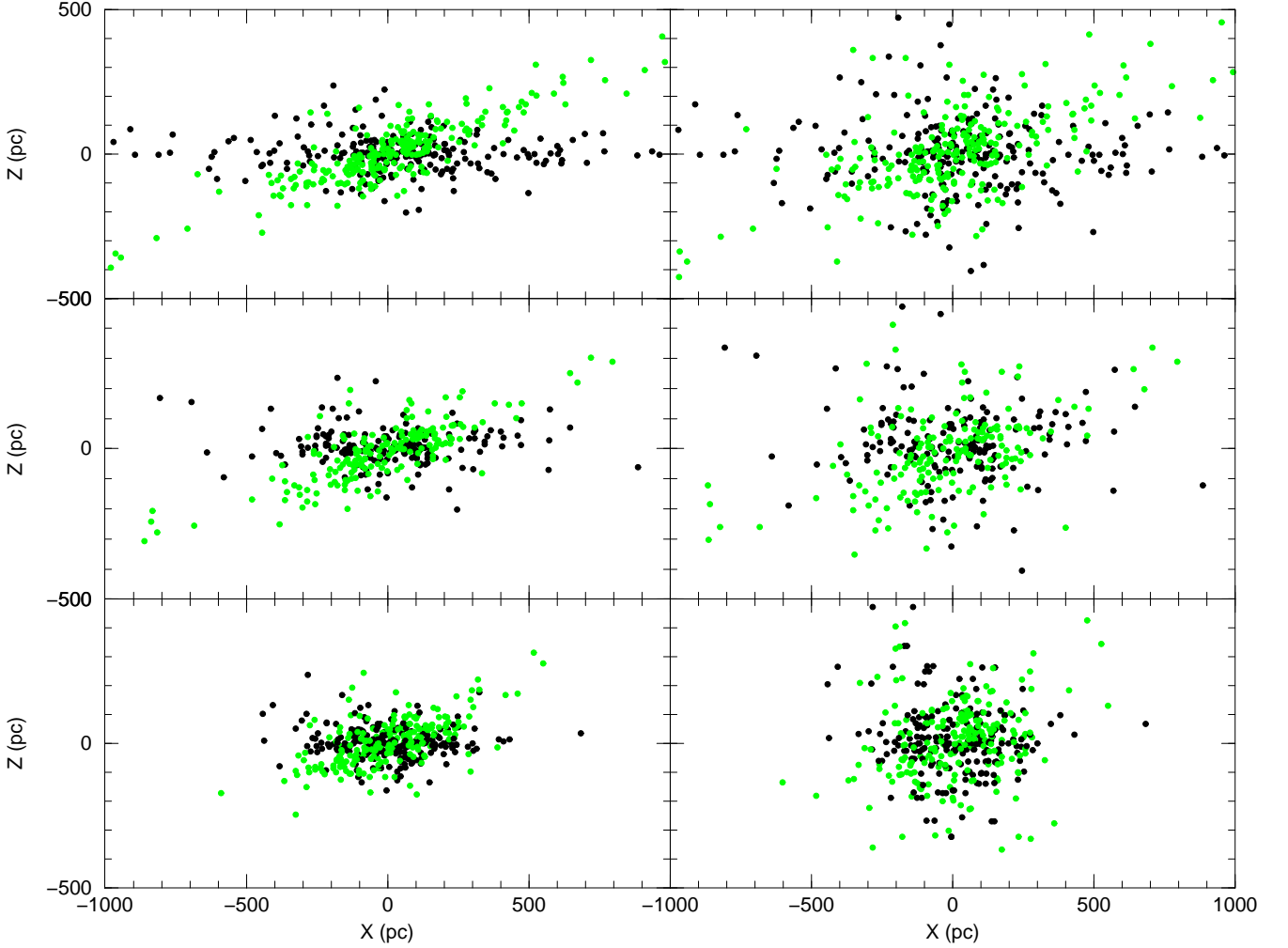}}
  \caption{Example of the distribution of the simulated samples on the $X$-$Z$ 
           galactic plane. At the top, the pseudo-stars with an age less
           than 30 Myr; in the middle, those with an age between 30 Myr
           and 60 Myr; and at the bottom, those with an age larger than 60
           Myr. Left: Gould and Galactic plane generated with $Z_0$ = 40
           pc; right: both generated with $Z_0$ = 80 pc.}
  \label{fig.sim.xz}
\end{figure*}

An example of the spatial distribution of the generated pseudo-stars is
presented in Fig. \ref{fig.sim.xz}, which can be compared with the
distribution of the real stars in Fig. \ref{fig.xz}. As seeked for, the
same incompleteness effects are present in our simulated samples. The
results after applying our resolution proces to the pseudo-stars with $V$
$\le$ 7 and $R$ $\le$ 600 pc are presented in Table \ref{tab.sim.Gould},
where in brackets we give the standard deviation for the 100 simulated
samples.

\begin{table}[top]
   \caption{Simulations on the Gould Belt's structural parameters. Results
            obtained after averaging 100 simulated samples with input
            values: $i_{\mathrm{G}}= 20\degr$, $\Omega_{\mathrm{G}} =
            270\degr$, $q = 0.50$, with scale height: 40, 60 and 80 pc. 
            Only pseudo-stars with $R \le$ 600 pc were considered.}
   \label{tab.sim.Gould}
\begin{tabular}{cccccr}
 \hline
$Z_0$&$i_{\mathrm{G}}$($\degr$)& $\Omega_{\mathrm{G}}$ ($\degr$) & $q$ & $\xi_{\mathrm{G}}$ ($\degr$) & $\xi_{\mathrm{g}}$ ($\degr$)\\
\hline
\hline
\multicolumn{6}{c}{$\tau \le$ 30 Myr} \\
\hline
\hline
\mbox{40} & 21.4$_{(3.0)}$ & 268.0$_{( 8.7)}$ & 0.50$_{(0.15)}$ & 14.5$_{(5.7)}$ & 15.7$_{(5.7)}$ \\
\mbox{60} & 22.7$_{(4.3)}$ & 268.5$_{(13.2)}$ & 0.52$_{(0.17)}$ & 19.3$_{(5.7)}$ & 18.7$_{(5.7)}$ \\
\mbox{80} & 21.6$_{(5.2)}$ & 268.5$_{(20.3)}$ & 0.51$_{(0.13)}$ & 24.2$_{(3.4)}$ & 24.2$_{(3.4)}$ \\
\hline
\hline
\multicolumn{6}{c}{30 $< \tau \le$ 60 Myr} \\
\hline
\hline
\mbox{40} & 21.1$_{(2.8)}$ & 269.7$_{( 9.4)}$ & 0.50$_{(0.13)}$ & 13.9$_{(4.6)}$ & 13.9$_{(4.0)}$ \\
\mbox{60} & 21.5$_{(4.2)}$ & 270.1$_{(15.9)}$ & 0.50$_{(0.14)}$ & 18.7$_{(4.6)}$ & 19.3$_{(4.6)}$\\
\mbox{80} & 22.3$_{(5.7)}$ & 268.1$_{(31.6)}$ & 0.49$_{(0.15)}$ & 22.3$_{(3.4)}$ & 23.0$_{(2.9)}$ \\
\hline
\hline
\multicolumn{6}{c}{60 $< \tau \le$ 90 Myr} \\
\hline
\hline
\mbox{40} & 21.1$_{(3.9)}$ & 269.0$_{(13.6)}$ & 0.50$_{(0.16)}$ & 15.1$_{(5.7)}$ & 15.7$_{(5.2)}$ \\
\mbox{60} & 22.8$_{(6.0)}$ & 269.3$_{(22.5)}$ & 0.49$_{(0.18)}$ & 19.3$_{(5.7)}$ & 20.5$_{(5.2)}$ \\
\mbox{80} & 22.8$_{(7.6)}$ & 262.7$_{(50.7)}$ & 0.50$_{(0.15)}$ & 24.8$_{(2.3)}$ & 24.8$_{(2.9)}$ \\
\hline
\hline
\multicolumn{6}{c}{90 $< \tau \le$ 120 Myr} \\
\hline
\hline
\mbox{40} & 20.9$_{(4.4)}$ & 271.3$_{(15.6)}$ & 0.49$_{(0.17)}$ & 15.1$_{(5.2)}$ & 16.3$_{(5.2)}$ \\
\mbox{60} & 21.7$_{(6.8)}$ & 269.1$_{(36.3)}$ & 0.48$_{(0.17)}$ & 20.5$_{(5.2)}$ & 21.1$_{(5.2)}$ \\
\mbox{80} & 24.7$_{(8.4)}$ & 253.2$_{(53.4)}$ & 0.50$_{(0.16)}$ & 25.5$_{(2.3)}$ & 25.5$_{(2.9)}$ \\
\hline
\end{tabular}
\end{table}

First, we confirm that the angular halfwidths ($\xi_{\mathrm{G}}$ and
$\xi_{\mathrm{g}}$) correctly reflect the growth of the scale height
($Z_0$) of the simulated belts, the standard deviation of the differents
samples ranging from $2\degr$ to $5\degr$. The $q$ parameter is also well
recovered ($q = 0.50$), though with a standard deviation as large as
0.13-0.17. On the other hand, although the values obtained for the angles
defining the Gould Belt's orientation ($i_{\mathrm{G}}$,
$\Omega_{\mathrm{G}}$) could indicate the presence of a small systematic
trend when increasing $Z_0$, probably due to the resolution process
applied (Comer\'on \cite{Comeron1}), it is always smaller than the
standard deviation quoted. Looking at the interval 90 $<$ $\tau$ $\le$ 120
Myr, where the observational incompleteness is more accentuated, we
realize that when the presence of the inclined structure (the Gould Belt)
is present in the simulated sample, only below the 2$\sigma$ level we
could obtain an $i_{\mathrm{G}}$ value as small as obtained from the real
sample ($i_{\mathrm{G}} = 3.4\degr$). That is, there is a probability less
than 5\% to obtain inclinations about $4\degr$ in the worst case ($Z_0 =
80$ pc).

%
\section{Simulations to check the kinematic analysis}\label{appendix2}

Numerical simulations allow us to quantitatively evaluate the biases in
the kinematic model parameters (Oort constants and solar motion
components) induced by both, our observational constraints -- irregular
spatial distribution of the stars, incompleteness effects, availability of
radial velocities, ... -- and the presence of observational errors in the
right hand side of Eqs. (\ref{eq.vr}), (\ref{eq.vl}) and (\ref{eq.vb}),
not considered in our least square fit. Here we present the procedure
followed to generate the simulated samples, the results derived when using
them to solve Eqs. (\ref{eq.vr}), (\ref{eq.vl}) and (\ref{eq.vb}), and
finally, the quantification of the biases present in our real sample
resolution.

\subsection{Process to generate the simulated samples}

To take into account the irregular spatial distribution of our stars and
their actual observational errors, parameters describing the position of
the each simulated pseudo-star were generated as follow:

\begin{itemize}

\item From each real star we generated a pseudo-star that has the same
nominal position ($R_0, l, b$) -- not affected by errors -- than the real
one.

\item We assumed that the angular coordinates ($l, b$) have negligible
observational errors.

\item If the star had a distance determination from the Hipparcos
parallax, the parallax error of the pseudo-star has a distribution law:

\begin{equation}\label{eq.dist.trig}
  \varepsilon(\pi) = e^{
  - \frac {1}{2} \left( \frac {\pi - \pi_0}{\sigma_\pi} \right)^2}
\end{equation}

\noindent where $\sigma_\pi$ is the individual error in the parallax
$\pi_0$ ($1/R_0$)  of the real star. From the value of $\pi$ affected by
the error, the simulated distance affected by error ($R = 1/\pi$) was
derived. On the other hand, if the star had a photometric distance
determination, the error follows:

\begin{equation}\label{eq.dist.fot}
  \varepsilon(R) = e^{
  - \frac {1}{2} \left( \frac {R - R_0}{\sigma_R} \right)^2}
\end{equation}

\noindent where $\sigma_R$ is the individual error in the photometric
distance of the star.

\end{itemize}

To generate kinematic parameters we randomly assigned to each pseudo-star
a velocity ($U, V, W$)  by assuming a cosmic dispersion ($\sigma_U$,
$\sigma_V$, $\sigma_W$) and a Schwarzschild distribution:
        
\begin{equation}
  \varphi'_v(U, V, W) = e^{
  - \frac {1}{2} \left( \frac {U - U_0}{\sigma_U} \right)^2
  - \frac {1}{2} \left( \frac {V - V_0}{\sigma_V} \right)^2
  - \frac {1}{2} \left( \frac {W - W_0}{\sigma_W} \right)^2}
\end{equation}

\noindent where ($U_0, V_0, W_0$) are the reflex of solar motion. These
components were transformed into radial velocities and proper motions in
galactic coordinates using the nominal position of the pseudo-star ($R_0,
l, b$). The systematic motion due to galactic rotation was added following
Eqs. (\ref{eq.vr}), (\ref{eq.vl}) and (\ref{eq.vb}), obtaining the
components ($v_{\mathrm{r_0}}, \mu_{\mathrm{l_0}}, \mu_{\mathrm{b_0}}$)
for each star. Finally, individual observational errors were introduced by
using the error function:

\begin{equation}
  \varepsilon(v_{\mathrm{r}}, \mu_{\mathrm{l}}, \mu_{\mathrm{b}}) = e^{
  - \frac {1}{2} \left( \frac {v_{\mathrm{r}} - v_{\mathrm{r_0}}}
  {\sigma_{v_{\mathrm{r}}}} \right)^2
  - \frac {1}{2} \left( \frac {\mu_{\mathrm{l}} - \mu_{\mathrm{l_0}}}
    {\sigma_{\mu_{\mathrm{l}}}} \right)^2
  - \frac {1}{2} \left( \frac {\mu_{\mathrm{b}} - \mu_{\mathrm{b_0}}}
    {\sigma_{\mu_{\mathrm{b}}}} \right)^2}
\end{equation}

\noindent where $\sigma_{v_{\mathrm{r}}}$, $\sigma_{\mu_{\mathrm{l}}}$ and
$\sigma_{\mu_{\mathrm{b}}}$ are the observational errors of the real star.

\begin{figure}
  \resizebox{\hsize}{!}{\includegraphics{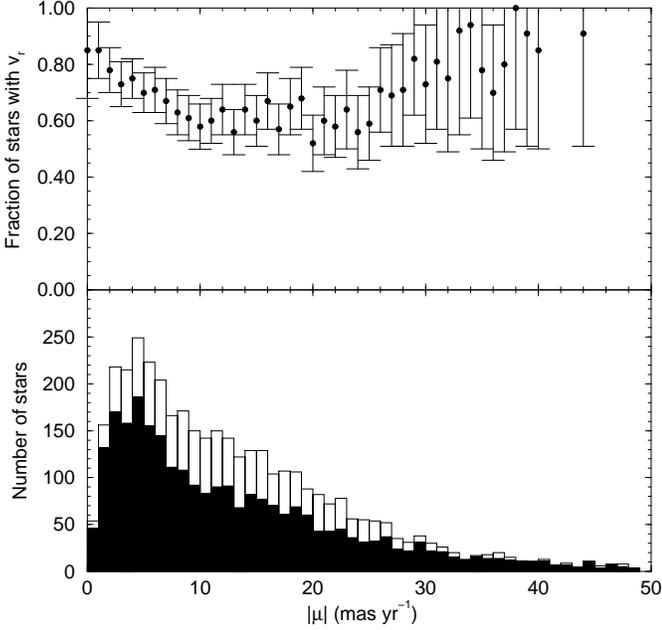}}
  \caption{Fraction of pseudo-stars with radial velocity (top) and distribution
           of the stars with proper motion (blank histogram) and radial 
           velocity (filled histogram) (bottom) as a function of proper motion
           for one of the simulated samples (3915 stars). Error
           bars were estimated from a Poissonian error distribution.}
  \label{fig.bias.vr.mu.sim}
\end{figure}

At the end of this process we had the following data for each pseudo-star:
galactic coordinates ($R,l,b$), velocity parameters ($v_{\mathrm{r}},
\mu_{\mathrm{l}}, \mu_{\mathrm{b}}$), errors in the velocity parameters
($\sigma_{v_{\mathrm{r}}}, \sigma_{\mu_{\mathrm{l}}},
\sigma_{\mu_{\mathrm{b}}}$) and error in the trigonometric parallax
($\sigma_\pi$) or in the photometric distance ($\sigma_R$). The simulated
radial component of those pseudo-stars generated from a real star without
radial velocity was not used, thus we imposed in the simulated sample the
same deficiency in radial velocity data that is present in our real sample
(Sect. \ref{vrad}). The fraction of pseudo-stars with radial velocity
against proper motion is shown in Fig. \ref{fig.bias.vr.mu.sim}. This can
be compared with Fig. \ref{fig.bias.vr.mu} in Sect. \ref{vrad} (the latter
was made using the whole catalogue (6922 stars) whereas the former only
contains the 3915 stars used in the fit of the Eqs. (\ref{eq.vr}),
(\ref{eq.vl}) and (\ref{eq.vb})). The systematic trend
present in the real sample is well reproduced in the simulations.

\subsection{Results and discussion}

Following this scheme, two sets of 100 simulated samples were built, each
one having the same number of stars than the real sample. The first set
was built adopting $K = 0$ km s$^{-1}$ kpc$^{-1}$ and used to derive the
kinematic parameters in the distance interval 600 $< R \le$ 2000 pc. For
the interval 100 $< R \le$ 600 pc we simulated the expansion by imposing
$K = 5$ km s$^{-1}$ kpc$^{-1}$. In Table \ref{tab.sim} we show the adopted
kinematic parameters. For the cosmic dispersion we considered
($\sigma_{\mathrm{U}}$, $\sigma_{\mathrm{V}}$, $\sigma_{\mathrm{W}}) =
(8,8,5)$ km s$^{-1}$ (see Sect. \ref{Oort}).

As can be seen in Table \ref{tab.sim}, three different resolution
processes were undertaken:

\begin{itemize}

\item {\bf Case 1}: A null error in distance was adopted. Therefore, the
nominal distance of the pseudo-star is used ($R = R_0$) and only the
effects of the errors on the radial velocity and proper motions are
considered. No stars are rejected.

\item {\bf Case 2}: Errors in the radial velocity, proper motions and
distance are considered. No stars are rejected.

\item {\bf Case 3}: As Case 2 but, to reproduce the real case, we rejected
those stars with a residual velocity 3 times larger than the root mean
square residual of the fit.

\end{itemize}

Case 1 allows to study the effect of the errors in radial velocity and
proper motions, the incompleteness of our sample, the lack of radial
velocity data (see Sect. \ref{vrad}) and the correlations between the
different kinematic parameters to be determined. As we can see in Table
\ref{tab.sim}, no systematic bias is present, the difference between the
adopted values and the obtained ones never exceeding 0.2 km s$^{-1}$ for
the solar motion components and 0.2 km s$^{-1}$ kpc$^{-1}$ for the Oort
constants.

When the error in the distance is considered (Case 2) the most noticeable
effect is a clear bias in the $A$ and $B$ Oort constants. For $A$ and the
radial velocity resolution, a bias of +(0.6-0.7) km s$^{-1}$ kpc$^{-1}$
was obtained. On the contrary, from the proper motion equations this bias
was $-$(0.7-0.8) km s$^{-1}$ kpc$^{-1}$. For the combined resolution, and
due to the larger number of proper motion equations ($2N$ against $N$),
the bias was $-$(0.1-0.5) km s$^{-1}$ kpc$^{-1}$, depending on the
distance interval considered. For $B$ constant, a bias of +(0.7-0.9) km
s$^{-1}$ kpc$^{-1}$ was found from proper motion data.  Apart from the
distance errors, other effects contribute to the biases detected: the
specific decreasing distribution on distance of our real sample (well
reproduced in the simulations) and the distance cut applied in the
resolution process (100, 600, 2000 pc).

In Case 3, to mimic the real case, those stars with a residual velocity
larger than 3 times the root mean square residual were rejected. In the
simulated samples there are not high velocity stars, but in the reality it
will be a few percentage of kinematically peculiar stars, or stars with
non-well determined errors in the distance or in the velocity components.
We confirm that our rejection criterion do not induced any addicional
bias, showing only an expected decrease in the $\chi^2$ statistic.

Concluding, these simulations allow us to estimate which biases are
expected in the kinematic parameters obtained from our real sample. As we
can see, for 100 $< R \le$ 600 pc our results for the combined solution
can be biased on $\approx -$0.5 km s$^{-1}$ kpc$^{-1}$ in $A$ Oort
constant and $\approx -$0.8 km s$^{-1}$ kpc$^{-1}$ in $B$, whereas for $C$
and $K$ is negligible. For the solar motion components a bias about
0.3-0.4 km s$^{-1}$ can be present. For 600 $< R \le$ 2000 pc the bias for
$A$, $C$ and $K$ is negligible while a positive bias of $\sim$ 0.9 km
s$^{-1}$ kpc$^{-1}$ is found for $B$. Again, for solar motion a bias of
0.3-0.4 km s$^{-1}$ is expected in each component.

\begin{table*}[top]
   \caption{Mean Oort constants and residual solar motion for 100
            simulated samples obtained soving Eq. (\ref{eq.vr}) for radial
            velocities, Eqs. (\ref{eq.vl}) + (\ref{eq.vb}) for proper 
            motions and Eqs. (\ref{eq.vr}) + (\ref{eq.vl}) + (\ref{eq.vb})
            for the combined solution. The standard deviation for the 100
            samples is shown in brackets. Units: $A$, $B$, $C$, $K$ in km
            s$^{-1}$ kpc$^{-1}$; $U_\odot$, $V_\odot$, $W_\odot$, $\sigma$ 
            in km s$^{-1}$. $\chi^2/N_{\mathrm{eq}}$ is the value of
            $\chi^2$ divided by the number of equations.} 
   \label{tab.sim}
\begin{tabular}{crrrrrrrr}
\hline
                          & \multicolumn{8}{c}{Radial velocities} \\
\hline
\hline
                          & \multicolumn{4}{c}{100 $< R \le$ 600 pc} & \multicolumn{4}{c}{600 $< R \le$ 2000 pc} \\
\hline
\hline
                          &\hspace{-0.5cm} Adopted & Case 1 & Case 2 & Case 3   
                          & Adopted & Case 1 & Case 2 & Case 3 \\
\hline
\hline
\hspace{-0.3cm}$A$        &\hspace{-0.5cm}    14.0 &   13.9$_{(1.8)}$ &   14.7$_{(1.6)}$ &    14.7$_{(1.7)}$
                          &    14.0 &   13.9$_{(0.7)}$ &   14.6$_{(0.8)}$ &    14.5$_{(0.8)}$ \\
\hspace{-0.3cm}$C$         &\hspace{-0.5cm}     0.0 &    0.2$_{(1.6)}$ & $-$0.1$_{(1.7)}$ &  $-$0.1$_{(1.7)}$  
                          &     0.0 & $-$0.1$_{(0.6)}$ &    0.1$_{(0.7)}$ &     0.1$_{(0.7)}$ \\
\hspace{-0.3cm}$K$        &\hspace{-0.5cm}    5.0 &    4.8$_{(1.1)}$ &     4.9$_{(1.2)}$ &     4.9$_{(1.2)}$ 
                          &     0.0 & $-$0.1$_{(0.4)}$ & $-$0.3$_{(0.9)}$ &  $-$0.3$_{(0.5)}$ \\
\hspace{-0.3cm}$U_\odot$  &\hspace{-0.5cm}     9.0 &    9.0$_{(0.5)}$ &    9.1$_{(0.4)}$ &     9.1$_{(0.5)}$ 
                          &     9.0 &    8.9$_{(0.8)}$ &    8.8$_{(0.9)}$ &     8.8$_{(0.9)}$ \\ 
\hspace{-0.3cm}$V_\odot$  &\hspace{-0.5cm}    12.0 &   12.1$_{(0.6)}$ &   12.2$_{(0.5)}$ &    12.2$_{(0.6)}$ 
                          &    12.0 &   12.1$_{(0.7)}$ &   11.6$_{(0.7)}$ &    11.6$_{(0.7)}$ \\
\hspace{-0.3cm}$W_\odot$  &\hspace{-0.5cm}     7.0 &    7.2$_{(1.0)}$ &    7.2$_{(1.0)}$ &     7.3$_{(1.0)}$    
                          &     7.0 &    7.0$_{(1.8)}$ &    6.7$_{(1.8)}$ &     6.7$_{(1.7)}$ \\
\hspace{-0.3cm}$\sigma$   &\hspace{-0.5cm}         &    8.4$_{(0.2)}$ &    8.5$_{(0.2)}$ &     8.4$_{(0.3)}$
                          &         &    8.3$_{(0.3)}$ &    9.2$_{(0.4)}$ &     9.0$_{(0.4)}$ \\
\hspace{-0.3cm}$\chi^2/N_{\mathrm{eq}}$  
                          &\hspace{-0.5cm}         &    1.00          &     1.00          &     0.97
                          &         &    1.00          &     1.11          &     1.06          \\
\hline
\hline
                          & \multicolumn{8}{c}{Proper motions} \\
\hline
\hline
                          & \multicolumn{4}{c}{100 $< R \le$ 600 pc} & \multicolumn{4}{c}{600 $< R \le$ 2000 pc} \\
\hline
\hline
                          &\hspace{-0.5cm} Adopted & Case 1 & Case 2 & Case 3 
                          & Adopted & Case 1 & Case 2 & Case 3 \\
\hline
\hline
\hspace{-0.3cm}$A$        &\hspace{-0.5cm}    14.0 &    14.2$_{(0.8)}$ &    13.2$_{(0.8)}$ &    13.2$_{(0.8)}$
                          &    14.0 &    14.0$_{(0.5)}$ &   13.3$_{(0.5)}$ &    13.3$_{(0.5)}$ \\
\hspace{-0.3cm}$B$        &\hspace{-0.5cm} $-$12.0 & $-$12.0$_{(0.7)}$ & $-$11.3$_{(0.7)}$ & $-$11.3$_{(0.7)}$ 
                          & $-$12.0 & $-$12.0$_{(0.4)}$ & $-$11.1$_{(0.4)}$ & $-$11.1$_{(0.4)}$ \\
\hspace{-0.3cm}$C$        &\hspace{-0.5cm}     0.0 &     0.1$_{(0.8)}$ &   $-$0.1$_{(0.8)}$ &  $-$0.1$_{(0.8)}$  
                          &     0.0 &     0.0$_{(0.5)}$ &  $-$0.1$_{(0.6)}$ &  $-$0.1$_{(0.6)}$ \\
\hspace{-0.3cm}$K$        &\hspace{-0.5cm}     5.0 &     5.2$_{(1.8)}$ &     4.8$_{(1.8)}$ &     4.8$_{(1.9)}$ 
                          &     0.0 &     0.1$_{(2.1)}$ &  $-$0.2$_{(2.1)}$ &  $-$0.3$_{(2.1)}$ \\
\hspace{-0.3cm}$U_\odot$  &\hspace{-0.5cm}     9.0 &     9.0$_{(0.2)}$ &    8.6$_{(0.2)}$ &     8.6$_{(0.2)}$ 
                          &     9.0 &     9.0$_{(0.6)}$ &     8.7$_{(0.7)}$ &     8.7$_{(0.7)}$ \\ 
\hspace{-0.3cm}$V_\odot$  &\hspace{-0.5cm}    12.0 &    12.0$_{(0.2)}$ &    11.4$_{(0.2)}$ &    11.4$_{(0.2)}$ 
                          &    12.0 &    12.1$_{(0.6)}$ &   11.4$_{(0.7)}$ &    11.4$_{(0.7)}$ \\
\hspace{-0.3cm}$W_\odot$  &\hspace{-0.5cm}     7.0 &     7.0$_{(0.1)}$ &     6.6$_{(0.1)}$ &     6.6$_{(0.1)}$    
                          &     7.0 &     7.1$_{(0.3)}$ &    6.7$_{(0.3)}$ &     6.7$_{(0.3)}$ \\
\hspace{-0.3cm}$\sigma$   &\hspace{-0.5cm}         &     6.4$_{(0.1)}$ &     6.6$_{(0.1)}$ &     6.6$_{(0.1)}$
                          &         &     7.6$_{(0.2)}$ &    7.8$_{(0.2)}$ &     7.6$_{(0.2)}$ \\
\hspace{-0.3cm}$\chi^2/N_{\mathrm{eq}}$  
                          &\hspace{-0.5cm}         &     1.00          &    0.94          &     0.91
                          &         &     1.00          &   0.91          &     0.88          \\
\hspace{-0.3cm}$(\chi^2/N_{\mathrm{eq}})_{\mathrm{l}}$ 
                          &\hspace{-0.5cm}         &     1.00          &     0.93          &     0.90
                          &         &     1.00          &   0.87          &     0.83          \\
\hspace{-0.3cm}$(\chi^2/N_{\mathrm{eq}})_{\mathrm{b}}$ 
                          &\hspace{-0.5cm}         &     1.00          &     0.94          &     0.91
                          &         &     1.00          &    0.96          &     0.93          \\
\hline
\hline
                          & \multicolumn{8}{c}{Combined solution} \\
\hline
\hline
                          & \multicolumn{4}{c}{100 $< R \le$ 600 pc} & \multicolumn{4}{c}{600 $< R \le$ 2000 pc} \\
\hline
\hline
                          &\hspace{-0.5cm} Adopted & Case 1 & Case 2 & Case 3 
                          & Adopted & Case 1 & Case 2 & Case 3 \\
\hline
\hline
\hspace{-0.3cm}$A$        &\hspace{-0.5cm}    14.0 &    14.1$_{(0.7)}$ &     13.5$_{(0.7)}$ &    13.5$_{(0.7)}$
                          &    14.0 &    14.0$_{(0.6)}$ &    13.9$_{(0.5)}$ &    13.9$_{(0.5)}$ \\
\hspace{-0.3cm}$B$        &\hspace{-0.5cm} $-$12.0 & $-$12.0$_{(0.7)}$ &  $-$11.2$_{(0.7)}$ & $-$11.2$_{(0.7)}$ 
                          & $-$12.0 & $-$12.0$_{(0.6)}$ & $-$11.1$_{(0.4)}$ & $-$11.2$_{(0.4)}$ \\
\hspace{-0.3cm}$C$        &\hspace{-0.5cm}     0.0 &     0.1$_{(0.7)}$ &      0.2$_{(0.7)}$ &     0.2$_{(0.7)}$  
                          &     0.0 &  $-$0.0$_{(0.5)}$ &     0.1$_{(0.4)}$ &     0.1$_{(0.4)}$ \\
\hspace{-0.3cm}$K$        &\hspace{-0.5cm}     5.0 &     5.0$_{(0.8)}$ &     5.1$_{(0.8)}$ &     5.1$_{(0.9)}$ 
                          &     0.0 &  $-$0.1$_{(0.5)}$ &  $-$0.2$_{(0.4)}$ &  $-$0.2$_{(0.4)}$ \\
\hspace{-0.3cm}$U_\odot$  &\hspace{-0.5cm}     9.0 &     9.0$_{(0.2)}$ &      8.7$_{(0.2)}$ &     8.7$_{(0.2)}$ 
                          &     9.0 &     9.0$_{(0.7)}$ &     8.8$_{(0.5)}$ &     8.7$_{(0.5)}$ \\ 
\hspace{-0.3cm}$V_\odot$  &\hspace{-0.5cm}    12.0 &    12.0$_{(0.2)}$ &     11.6$_{(0.2)}$ &    11.6$_{(0.2)}$ 
                          &    12.0 &    12.1$_{(0.6)}$ &    11.6$_{(0.5)}$ &    11.6$_{(0.5)}$ \\
\hspace{-0.3cm}$W_\odot$  &\hspace{-0.5cm}     7.0 &     7.0$_{(0.1)}$ &      6.7$_{(0.1)}$ &     6.7$_{(0.1)}$    
                          &     7.0 &     7.1$_{(0.4)}$ &     6.7$_{(0.3)}$ &     6.7$_{(0.3)}$ \\
\hspace{-0.3cm}$\sigma$   &\hspace{-0.5cm}         &     6.7$_{(0.1)}$ &      6.7$_{(0.1)}$ &     6.6$_{(0.1)}$
                          &         &    10.2$_{(0.2)}$ &     8.1$_{(0.2)}$ &     8.0$_{(0.2)}$ \\
\hspace{-0.3cm}$\chi^2/N_{\mathrm{eq}}$  
                          &\hspace{-0.5cm}         &     1.00          &     0.95          &     0.92
                          &         &     1.00          &      0.97          &     0.93          \\
\hspace{-0.3cm}$(\chi^2/N_{\mathrm{eq}})_{\mathrm{r}}$ 
                          &\hspace{-0.5cm}         &     1.00          &     1.00          &     0.97   
                          &         &     1.00          &      1.12          &     1.08          \\
\hspace{-0.3cm}$(\chi^2/N_{\mathrm{eq}})_{\mathrm{l}}$ 
                          &\hspace{-0.5cm}         &     1.00          &     0.93          &     0.90
                          &         &     1.00          &      0.98          &     0.84          \\
\hspace{-0.3cm}$(\chi^2/N_{\mathrm{eq}})_{\mathrm{b}}$ 
                          &\hspace{-0.5cm}         &     1.00          &     0.94          &     0.91
                          &         &     1.00          &      0.96          &     0.93          \\
\hline
\end{tabular}
\end{table*}

\end{document}